\documentclass[reprint,aip,rsi,amsmath,amssymb,nofootinbib]{revtex4-1}

\usepackage[normalem]{ulem}
\usepackage{graphicx}
\usepackage{color,soul}
\usepackage{subfigure}
\usepackage{dcolumn}
\usepackage{bm}
\usepackage{hyperref}
\usepackage{url}
\usepackage{multirow}
\hypersetup{
    colorlinks=true,
    linkcolor=blue,
    filecolor=blue,      
    urlcolor=blue,
    citecolor=blue,
}
\usepackage{float}

\newcommand{\etc}{\textit{etc.}}
\newcommand{\ie}{\textit{i.e. }}
\newcommand{\eg}{\textit{e.g., }}
\newcommand{\vs}{\textit{vs. }}

\usepackage{longtable}

\begin{document}

\title{Monte Carlo Modeling of Spin-polarized Photoemission from $p$-doped GaAs Activated to Negative Electron Affinity}

\author{Oksana Chubenko}
\email[E-mail address: ]{chubenko@asu.edu}
\affiliation{Department of Physics, Arizona State University, Tempe, AZ 85287, USA}

\author{Siddharth Karkare}
\affiliation{Department of Physics, Arizona State University, Tempe, AZ 85287, USA}

\author{Dimitre Dimitrov}
\affiliation{Los Alamos National Laboratory, Santa Fe, NM 87545, USA}

\author{Jai Kwan Bae}
\affiliation{Department of Physics, Cornell University, Ithaca, NY 14853, USA}

\author{Luca Cultrera}
\affiliation{Department of Physics, Cornell University, Ithaca, NY 14853, USA}

\author{Ivan Bazarov}
\affiliation{Department of Physics, Cornell University, Ithaca, NY 14853, USA}

\author{Andrei Afanasev}
\affiliation{Department of Physics, The George Washington
University, Washington, DC 20052, USA}

\begin{abstract}
The anticorrelation between quantum efficiency (QE) and electron spin polarization (ESP) from a $p$-doped GaAs activated to negative electron affinity (NEA) is studied in detail using an ensemble Monte Carlo approach. The photoabsorption, momentum and spin relaxation during transport, and tunnelling of electrons through the surface potential barrier are modeled to identify fundamental mechanisms, which limit the efficiency of GaAs spin-polarized electron sources. In particular, we study the response of QE and ESP to various parameters such as the photoexcitation energy, doping density, and electron affinity level. Our modeling results for various transport and emission characteristics are in a good agreement with available experimental data. Our findings show that the behaviour of both QE and ESP at room temperature can be fully explained by the bulk relaxation mechanisms and the time which electrons spend in the material before being emitted. 
\end{abstract}
\maketitle

\section{\label{sec:introduction}Introduction}
GaAs and GaAs-based photocathodes activated to negative electron affinity (NEA) is the only existing technology to produce spin-polarized electron beams of high intensities for advanced nuclear-physics and particle-physics experiments designed to study nucleon spin structures, parity-violating mechanisms, and other spin-dependent phenomena at leading accelerator facilities, including SLAC and Jefferson Lab's Continuous Electron Beam Accelerator Facility (CEBAF).\cite{Hernandez_2008} These experiments set a number of requirements for electron sources, including high quantum efficiency $QE$ and high electron spin polarization $ESP$ to maximize the expression $ESP^2\cdot QE$, which is often used\cite{Pierce_1980,Erbudak_1978}  as a figure of merit when discussing efficiency of different spin-polarized photocathodes.

Spin-polarized electron beams are expected to play an even more important role in future experiments at BNL's Electron-Ion Collider (EIC). Those experiments will require at least 85$\%$ initial electron beam polarization and the collision luminosity as high as $10^{34}$ cm$^{-2}$s$^{-1}$. Therefore, the pursuit for more robust photocathodes with higher QE, higher ESP, lower emittance, and faster response times is an active research field nowadays.\cite{Tsentalovich_2019, Liu_2016}

Characteristically, the fabrication methods and operational conditions commonly accepted to increase QE (\eg high doping concentrations, high photoexcitation energies, negative electron affinity levels, \etc) lead to spin depolarization, and vice versa.\cite{Liu_2017,Mccarter_2010,Erbudak_1978} Therefore, the availability of effective modeling tools which can be used to enhance our understanding of processes which affect the performance of conventional spin-polarized electron sources and identify novel candidates with potentially improved capabilities is of utmost importance. Such tools are essential to design advanced photocathodes and define operational conditions that can deliver stringent demands of high spin polarization and high beam currents simultaneously as required by various applications.

The fundamental three-step photoemission model for semiconductors was developed by Spicer\cite{Spicer_1958} in the 1950s for alkali antimonide photocathodes. More practical formalization of Spicer's model represented by a simple parametric expression was developed later\cite{Spicer_1977} for NEA III-V compounds. Being based on the thermal diffusion equation for  electron transport, the early models provide a good approximation as long as the assumption of bulk optical absorption (\ie photoexcitation with infrared light) works.\cite{Spicer_1977, Lundstrom_Fundamentals} In cases when the bulk-absorption assumption does not hold true, the Monte Carlo approach, a numerical technique for solving the Boltzmann Transport Equation, can be used to simulate electron transport in photocathodes.

Over the last several decades, significant attempts were made to develop a detailed Monte Carlo model for electron transport in bulk semiconductor materials and devices. \cite{Jacoboni_1983,Tomizawa_Numerical,Vasileska_2010} In a combination with photoexcitation and emission into the vacuum, the Monte Carlo approach has proven its reliability and convenience in modeling unpolarized photoemission from bulk and layered GaAs, showing a good agreement with experimental measurements.\cite{Karkare_2013,Karkare_2014,Liu_2019} As for the spin polarization, the spin-relaxation mechanisms\cite{Elliott_1954,Yafet_1963,D'yakonov_1972,Bir_1976} in GaAs photocathodes have been widely studied\cite{Fishman_1977,Aronov_1983,Zerrouati_1988, Song_2002,Dyson_2004} and the spin relaxation during electron transport in bulk GaAs has been investigated. \cite{Sanada_2002, Barry_2003,Jiang_2009} However, the complete from-excitation-to-emission-into-vacuum photoemission of spin-polarized electrons has not yet been implemented in numerical simulations. 

In the present work, we modify the Monte Carlo transport model\cite{Vasileska_2010} to include the photoexcitation and emission of electrons. We also implement depolarization mechanisms to calculate QE and ESP simultaneously for doping densities between $5\times 10^{17}$ and $10^{19}$ cm$^{-3}$ in the photoexcitation energy range from the band-gap energy to 2.2~eV with only the electron affinity used as a free fitting parameter. 

This paper is organized as follows. General description of the three-step model of spin-polarized electron photoemission and the band-structure model are given in Section~\ref{sec:three-step model}. Section~\ref{sec:implementation} provides physics and implementation details of the photoexcitation, spin-polarized electron transport, and the main surface effects. In Section~\ref{sec:results}, we investigate the influence of the electron affinity level and doping density on the QE and ESP, and compare our results to the experimental data available in literature. Section~\ref{sec:conclusions} summarizes our results and findings and highlights possible future directions for applying the developed model.

\section{\label{sec:three-step model}Three-Step Model of Spin-polarized Photoemission from Semiconductors}
\begin{figure}[b]
\centering
\includegraphics[width=3in]{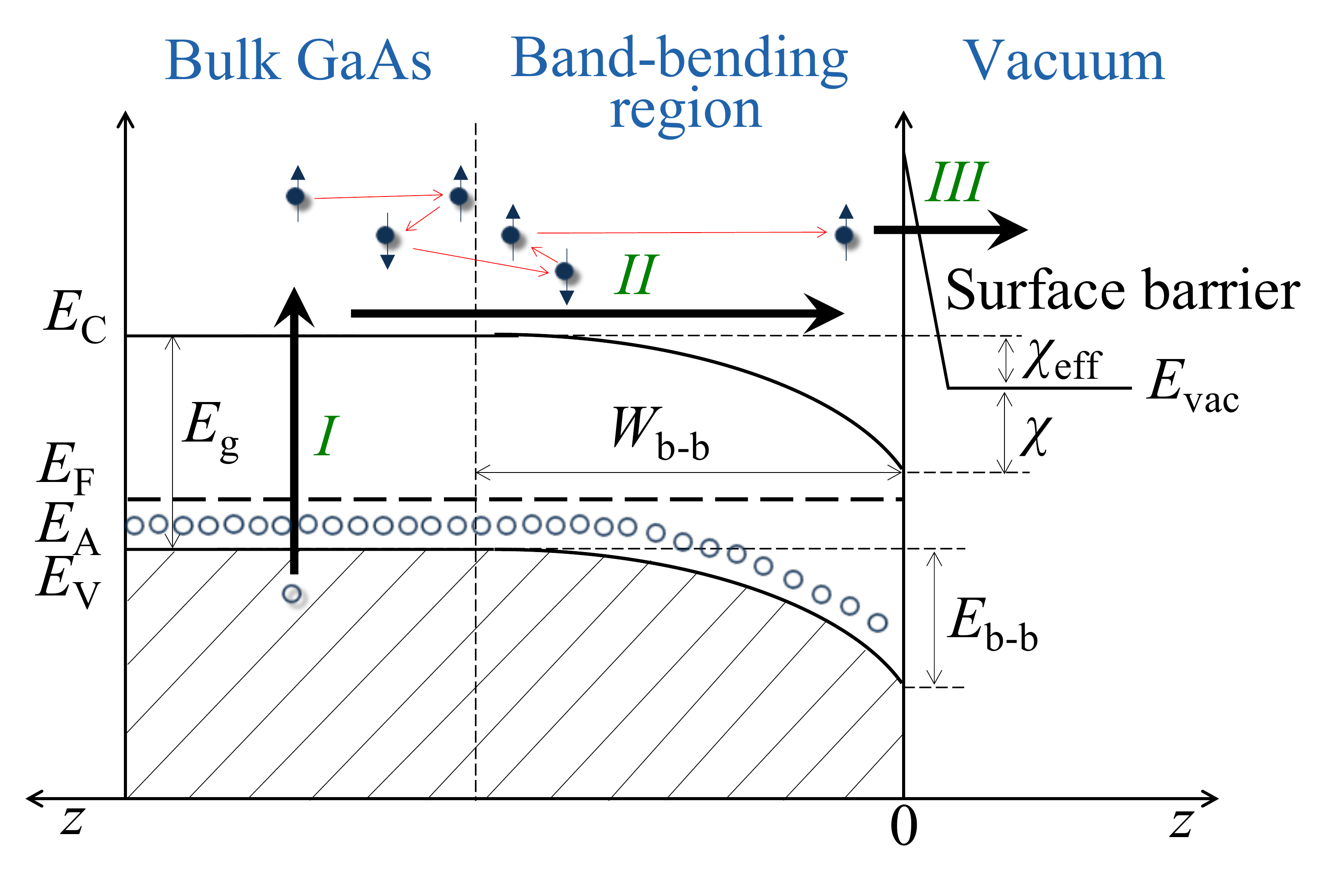}
\caption{Three-step model of spin-polarized photoemission from $p$-type NEA GaAs: I -- photoexcitation; II -- transport; III -- emission into the vacuum.}
\label{GaAs_model}
\end{figure}

The photoemission from semiconductors occurs in three steps: photoexcitation of electrons from the valence band (VB) to the conduction band (CB), transport to the surface, and emission into the vacuum, as shown in Fig.~\ref{GaAs_model}. Here $E_\text{C}$ and $E_\text{V}$ define the conduction band minimum (CBM) and the valence band maximum (VBM), respectively. $E_\text{F}$ and $E_\text{A}$ are the Fermi level and the acceptors level, respectively, and $E_\text{g}$ is the band gap. The band-bending region, which occurs at the surface of heavily $p$-doped semiconductors, is characterized by the band-bending width $W_{\text{b-b}}$ and the band-bending depth $E_{\text{b-b}}$. The position of the vacuum level $E_{\text{vac}}$ with respect to the CBM at the surface and in the bulk is defined by the electron affinity $\chi$ and the effective electron affinity $\chi_\text{eff}$, respectively.

\begin{figure}[!t]
\centering
\includegraphics[width=3.2in]{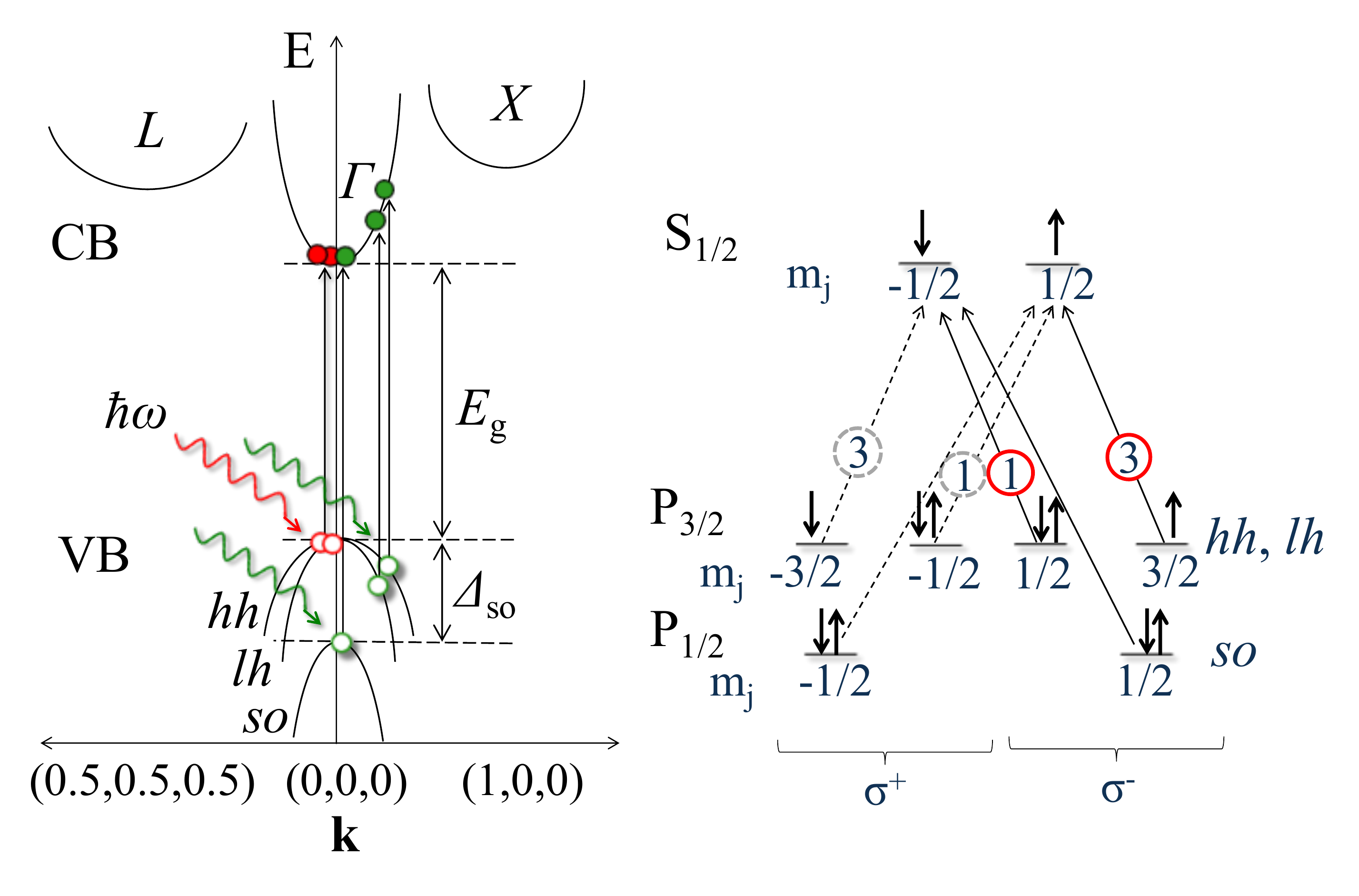}
\caption{Electromagnetic transitions at the $\Gamma$ point of unstrained GaAs under the excitation by circularly polarized light (right-polarized $\sigma^+$ or left-polarized $\sigma^-$). Due to the optical selection rules, only the transition for which $\Delta m_j=m_f-m_i=+1$ for $\sigma^+$ (or $\Delta m_j=-1$ for $\sigma^-$) are possible. Different colors correspond to excitations by photons of different energies. Numbers inside the circles indicate the relative strength of transitions under the photoexcitation by nearly band-gap-energy photons. Arrows $\uparrow$ and $\downarrow$ indicate the spins parallel and antiparallel to the direction of light propagation.}
\label{GaAs_structure}
\end{figure}

The band-structure model for GaAs is shown in Fig.~\ref{GaAs_structure}. The first conduction band has three minima called $\Gamma$ valley (coincides with the center of the Brillouin zone), L valley, and X valley. Each CB valley can be described by the dispersion relation
\begin{equation}
E_{\bold k}(1+\alpha E_{\bold k})=\frac{\hbar^2 k^2}{2m^*_\text{e}}\equiv\gamma_{\bold k},
\label{disp_CB}
\end{equation}
or, alternatively,
\begin{equation}
E_{\bold k} = \frac{\sqrt{1+4\alpha\hbar^2 k^2/(2m_\text{e}^*)}-1}{2\alpha},
\end{equation}
where $E_{\bold k}$ is the electron energy with respect to the bottom of a particular valley in a state with the wave vector $\bold{k}$, $m^*_\text{e}$ is the effective mass of the electron in a particular valley, $\alpha$ is the non-parabolicity factor, and $\hbar$ is the reduced Planck's constant. The structure of a VB is represented by the heavy hole ($hh$), light hole ($lh$), and split-off ($so$) sub-bands. Each of them has a maximum at $\bold{k}=0$ and can be described by
\begin{equation}
E_{\bold k}=-\frac{\hbar^2 k^2}{2m^*_\text{h}},
\label{disp_VB}
\end{equation} 
where $E_{\bold k}$ is the energy of the hole with respect to the maximum of a particular VB and $m^*_\text{h}$ is the effective mass of the hole.

Near the center of the Brillouin zone, where $\bold k$ is zero, direct transitions between the P-type ($l$ = 1) VB and S-type ($l$ = 0) CB states  are possible. The CB state is doubly degenerate ($j=1/2$; $m_j=\pm 1/2$). As for the VB, a four-fold degenerate state ($j=3/2$; $m_j=\pm 3/2$, $\pm 1/2$) is separated from a doubly degenerate state ($j=1/2$; $m_j=\pm 1/2$) by an energy distance $\Delta_{\text{so}}\approx$ 0.3 eV, the spin-orbit splitting. Here the quantum numbers $l$, $j$, and $m_j$ are the orbital angular momentum, total angular momentum, and the projection of total angular momentum onto the $z$ axis taken along the direction of the light propagation, respectively.

Both ESP and QE depend on the photon energy $\hbar \omega$. When an unstrained GaAs is illuminated by circularly polarized light with the photon energy in the range $E_\text{g} < \hbar \omega < E_\text{g}+\Delta_{\text{so}}$, two transitions from the $P_{3/2}$ state are allowed, with three times as many electrons in one spin state as in the other spin state (see Fig.~\ref{GaAs_structure}). So theoretically, an unstrained GaAs can provide maximum 50$\%$ spin polarization $ESP_0$ under  photoexcitation by circularly polarized light with $\hbar \omega \approx E_\text{g}$
\begin{equation}
ESP_0\equiv \frac{N_{\uparrow}-N_{\downarrow}}{N_{\uparrow}+N_{\downarrow}}=\frac{3-1}{3+1}=0.5.
\end{equation}
However, due to different depolarizing mechanisms in the material, the experimentally measured polarization is limited at about 35$\%$. \cite{Hernandez_2008} As the photon energy increases, photoexcitations from all three valence sub-bands become possible, reducing the polarization to zero.

In the matter of quantum efficiency $QE$, which is defined as a ratio of the number of emitted electrons $N_{\text{e}^-}$ to the number of incident photons $N_{\gamma}$
\begin{equation}
QE\equiv\frac{N_{\text{e}^-}}{N_{\gamma}},
\end{equation}
it continuously increases with photon energy since electrons gain more energy and have higher probability to reach the surface and escape into the vacuum. 

Below, we provide implementation details of three steps of photoemission from NEA GaAs as well as a description of physical processes and mechanisms implemented in the model. The room-temperature parameters used in the calculations are listed in Table~\ref{GaAs_parameters}.

\begin{table}[t]
\caption{Material parameters of GaAs used in the calculations ($T=300$ K).}
\resizebox{\linewidth}{!}{
\begin{tabular}{c|c|c}
\hline
\hline
Symbol & Meaning, units & Value [Ref.] \\
\hline
\multicolumn{2}{l}{Band model parameters} \\
$m_{\Gamma}^*$ & Electron effective mass in $\Gamma$ valley, $m_0$ & 0.063 [\onlinecite{Blakemore_1982}]\\
$m_{L}^*$ & Electron effective mass in $L$ valley, $m_0$ & 0.22 [\onlinecite{Lundstrom_Fundamentals}]\\
$m_{X}^*$ & Electron effective mass in $X$ valley, $m_0$ & 0.58 [\onlinecite{Lundstrom_Fundamentals}]\\
$\alpha_{\Gamma}$ & Non-parabolicity factor for $\Gamma$ valley, eV$^{-1}$ & 0.61 [\onlinecite{Lundstrom_Fundamentals}]\\
$\alpha_{L}$ & Non-parabolicity factor for $L$ valley, eV$^{-1}$ & 0.461 [\onlinecite{Lundstrom_Fundamentals}]\\
$\alpha_{X}$ & Non-parabolicity factor for $X$ valley, eV$^{-1}$ & 0.204 [\onlinecite{Lundstrom_Fundamentals}]\\
$m_{hh}^*$ & $hh$ effective mass, $m_0$ & 0.50 [\onlinecite{Blakemore_1982}]\\
$m_{lh}^*$ & $lh$ effective mass, $m_0$ & 0.088 [\onlinecite{Blakemore_1982}]\\
$m_{so}^*$ & $so$ effective mass, $m_0$ & 0.15 [\onlinecite{Zollner_2001}]\\
$E_{g0}$ & Intrinsic band gap energy, eV & 1.423 [\onlinecite{Blakemore_1982}] \\
$\Delta_{so}$ & Split-off energy gap, eV & 0.332 [\onlinecite{Zollner_2001}] \\
$\Delta_{\Gamma L}$ & Energy splitting between minima of $\Gamma$ and $L$ valleys, eV & 0.284 [\onlinecite{Blakemore_1982}]\\
$\Delta_{\Gamma X}$ & Energy splitting between minima of $\Gamma$ and $X$ valleys, eV  &  0.476 [\onlinecite{Blakemore_1982}] \\
\multicolumn{2}{l}{Momentum relaxation parameters} \\
$\Xi_{d\Gamma}$ & Acoustic deformation potential for $\Gamma$ valley, eV  & 7.01 [\onlinecite{Lundstrom_Fundamentals}]\\
$\Xi_{d L}$ & Acoustic deformation potential for $L$ valley, eV  & 9.2 [\onlinecite{Lundstrom_Fundamentals}]\\
$\Xi_{d X}$ & Acoustic deformation potential for $X$ valley, eV  & 9.0 [\onlinecite{Lundstrom_Fundamentals}]\\
$\hbar \omega_0$ & Polar optical phonon energy, meV  & 35.36 [\onlinecite{Lundstrom_Fundamentals}]\\
$D_{\Gamma L}$ & Deformation potential for $\Gamma\to L$ scattering, eV \AA$^{-1}$  & 10 [\onlinecite{Lundstrom_Fundamentals}]\\
$D_{\Gamma X}$ & Deformation potential for $\Gamma\to X$ scattering, eV \AA$^{-1}$  & 10 [\onlinecite{Lundstrom_Fundamentals}]  \\
$D_{L L}$ & Deformation potential for $L\to L$ scattering, eV \AA$^{-1}$  & 10 [\onlinecite{Lundstrom_Fundamentals}] \\
$D_{L X}$ & Deformation potential for $L\to X$ scattering, eV \AA$^{-1}$  & 5 [\onlinecite{Lundstrom_Fundamentals}] \\
$D_{X X}$ & Deformation potential for $X\to X$ scattering, eV \AA$^{-1}$  & 7 [\onlinecite{Lundstrom_Fundamentals}] \\
$\hbar \omega_{\Gamma L}$ & Intervalley phonon energy for $\Gamma\to L$ scattering, meV  & 27.8 [\onlinecite{Lundstrom_Fundamentals}] \\
$\hbar \omega_{\Gamma X}$ & Intervalley phonon energy for $\Gamma\to X$ scattering, meV  & 29.9 [\onlinecite{Lundstrom_Fundamentals}] \\
$\hbar \omega_{L L}$ & Intervalley phonon energy for $L\to L$ scattering, meV  & 29 [\onlinecite{Lundstrom_Fundamentals}] \\
$\hbar \omega_{L X}$ & Intervalley phonon energy for $L\to X$ scattering, meV  & 29.3 [\onlinecite{Lundstrom_Fundamentals}] \\
$\hbar \omega_{X X}$ & Intervalley phonon energy for $X\to X$ scattering, meV  & 29.9 [\onlinecite{Lundstrom_Fundamentals}] \\
$Z_{\Gamma}$ & Number of equivalent $\Gamma$ valleys to scatter into & 1 [\onlinecite{Vasileska_Computational}] \\
$Z_{L}$ & Number of equivalent $L$ valleys to scatter into & 4 [\onlinecite{Vasileska_Computational}] \\
$Z_{X}$ & Number of equivalent $X$ valleys to scatter into & 3 [\onlinecite{Vasileska_Computational}] \\
\multicolumn{2}{l}{Spin relaxation parameters} \\
$A_\text{ap}$ &   EY constant for scattering by acoustic phonons & 32/27 [\onlinecite{Fishman_1977}]   \\
$A_\text{pop}$ &   EY constant for scattering by polar optical phonons & 32/27 [\onlinecite{Fishman_1977}]   \\
$A_{ij}$ &   EY constant for intervalley scatterings & 32/27 [\onlinecite{Fishman_1977}]   \\
$A_\text{ii}$ & EY constant for scattering by ionized impurities  & 32/27 [\onlinecite{Fishman_1977}]   \\
$Q_\text{ap}$ &   DP constant for scattering by acoustic phonons & 1/6 [\onlinecite{Fishman_1977}]   \\
$Q_\text{pop}$ &   DP constant for scattering by polar optical phonons & 1/6 [\onlinecite{Fishman_1977}]   \\
$Q_{ij}$ & DP constant for intervalley scatterings  & 1/6 [\onlinecite{Fishman_1977}]   \\
$Q_\text{ii}$ & DP constant for scattering by ionized impurities  & 1/6 [\onlinecite{Fishman_1977}]   \\
$\Delta_\text{exc}$ & Exchange splitting of exciton ground state, $\mu$eV & 47 [\onlinecite{Aronov_1983}]\\
$\big|\psi(0)\big|^2$ & Sommerfeld factor & 1 [\onlinecite{Jiang_2009}]\\
\multicolumn{2}{l}{Other material parameters} \\
$\epsilon_{\infty}$ & High-frequency dielectric constant, $\epsilon_0$ &  10.92 [\onlinecite{Lundstrom_Fundamentals}]\\
$\epsilon_s$ &  Static dielectric constant, $\epsilon_0$&12.90 [\onlinecite{Lundstrom_Fundamentals}] \\
$\rho$ & Crystal density, kg m$^{-3}$ & 5360 [\onlinecite{Lundstrom_Fundamentals}] \\
$v_s$ & Sound velocity, m s$^{-1}$ & 5240 [\onlinecite{Lundstrom_Fundamentals}] \\
\hline
\hline
\end{tabular}
}
\label{GaAs_parameters}
\end{table}

\section{\label{sec:implementation}Implementation of the Model}
\subsection{\label{sec:photoexcitation}Photoexcitation}

\begin{figure}[!t]
\centering
\includegraphics[width=3in]{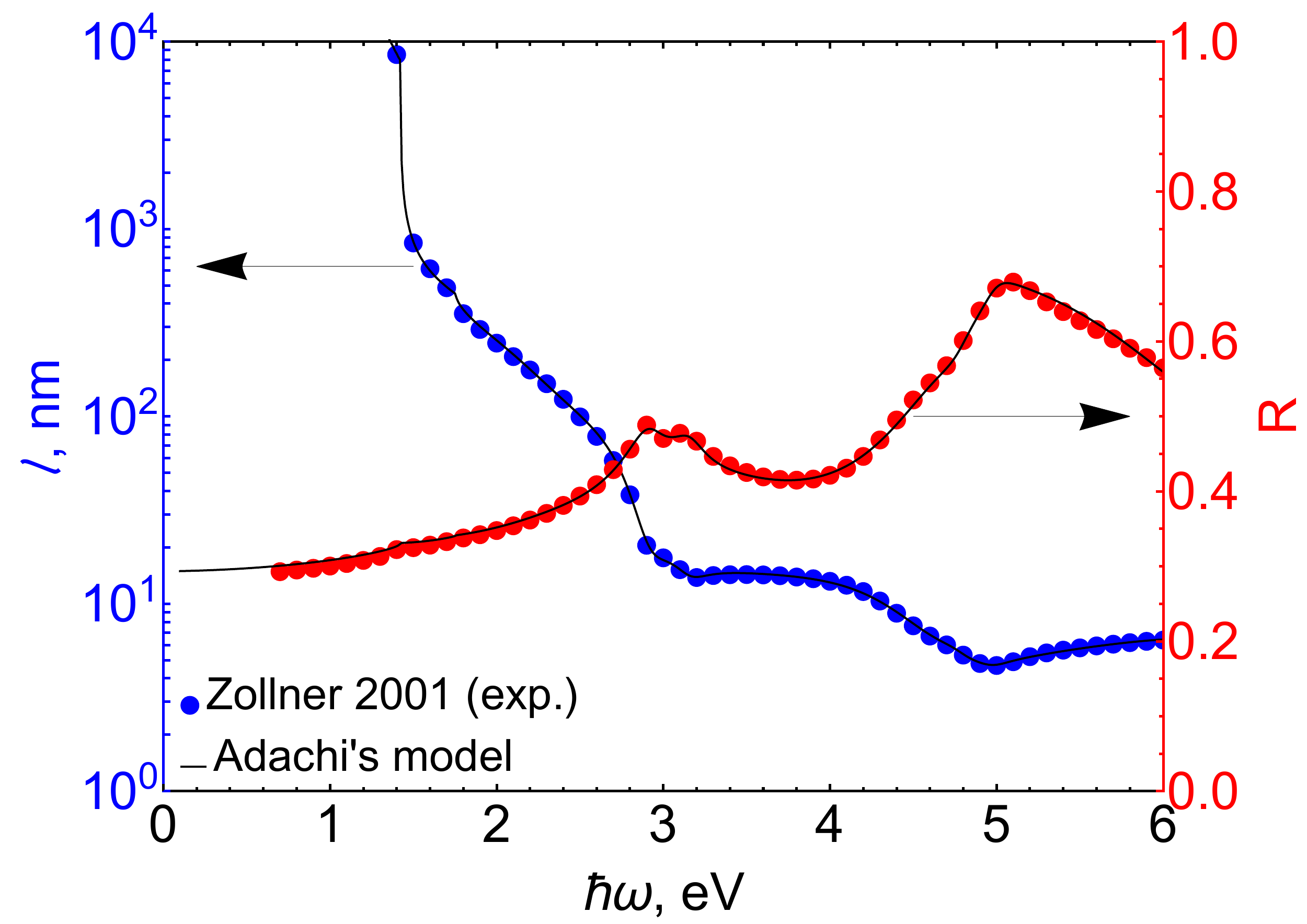}
\caption{GaAs absorption length $l$ and reflectivity $R$ as a function of the photoexcitation energy $\hbar\omega$.}
\label{alpha_R}
\end{figure}

The simulation begins with generation of $N_{\gamma 0}$ photons incident along the $z$ axis perpendicular to the GaAs surface. The generation rate of electron-hole pairs in the material is given by \cite{Vergara_1997}
\begin{equation}
\varrho(z) = I_0 (1-R)  \exp(-z/l) / l,
\label{g}
\end{equation}
where $I_0$ is the intensity of incident light, $R$ is the optical reflection coefficient, and $l$ is the absorption length. We assume illumination of heavily $p$-doped GaAs by low-intensity light, so the concentration of nonequilibrium carriers created during the photoexcitation is small in comparison with the concentration of equilibrium holes. For simplicity, we assume that the laser pulse is a delta function in time.\cite{Osman_1987} Also, we neglect any multielectron and multiphoton events and consider only the processes in which one photon interacts with one electron and photoexcites it directly to the $\Gamma$ valley. Optical parameters $R$ and $l$ depend on the photoexcitation energy and can be found from fitting the experimental data with Adachi's model\cite{Ozaki_1995} of dielectric function. In Fig.~\ref{alpha_R}, we show the resulting $l$ \vs $\hbar \omega$ and $R$ \vs $\hbar \omega$ curves. The experimental data are taken from Ref.~\onlinecite{Zollner_2001}.

\begin{figure}[!b]
\centering
\includegraphics[width=3in]{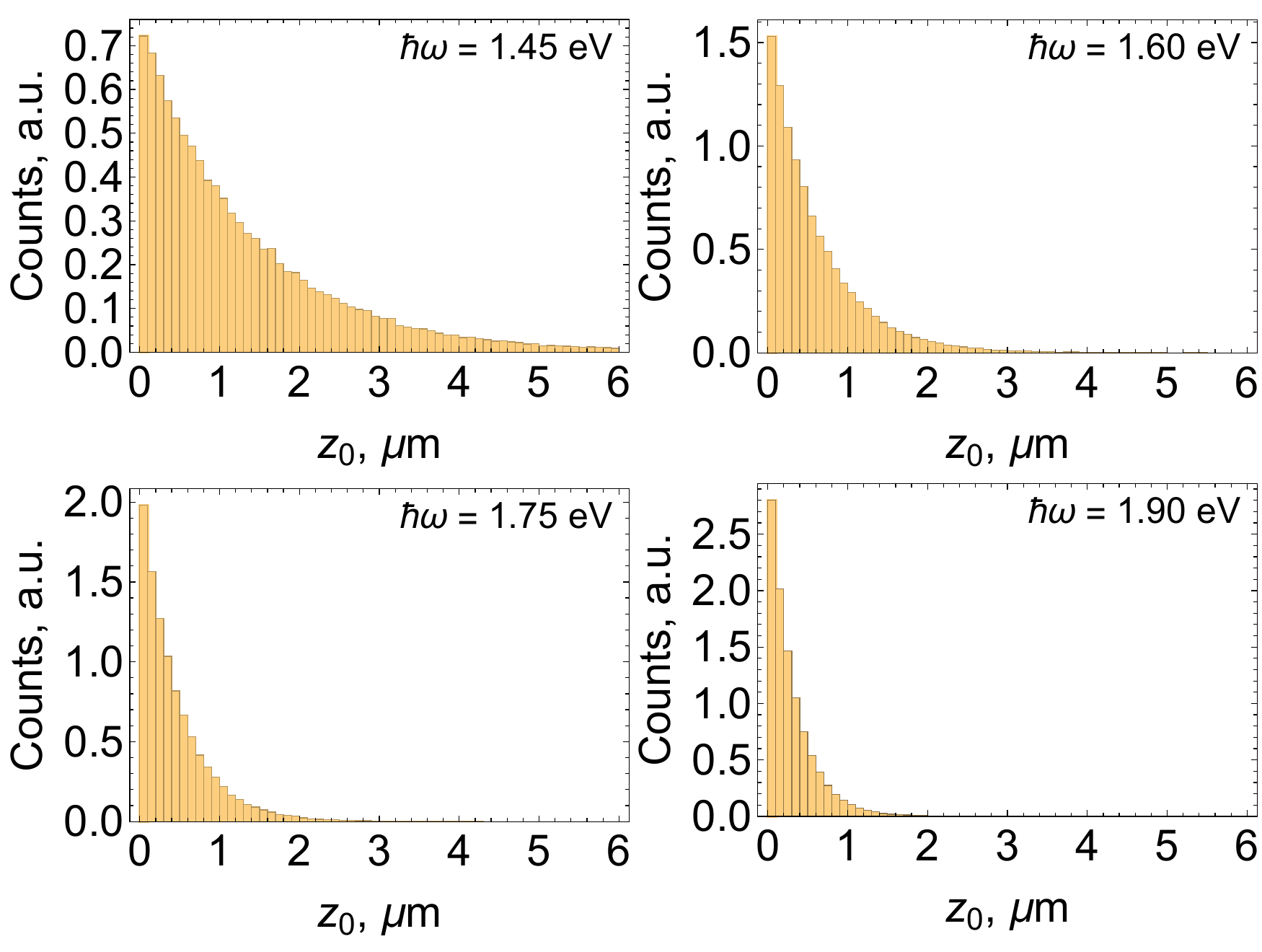}
\caption{Initial $z$ coordinates of photoexcited electrons for different photon energies. The bin width of the histograms is equal to 100~nm.}
\label{InCoord}
\end{figure}

Assuming an infinitely thick sample, the initial $z$ coordinate of a photoexcited electron can be generated using a random number $r$ between zero and unity as\cite{Joshi_1990}
\begin{equation}
z_0=-l\ln(1-r).
\end{equation}
The resulting histograms for different photoexcitation energies are shown in Fig.~\ref{InCoord}. 1.45 eV photons penetrate as deep as 6~$\mu$m into the material, whereas high-energy photons excite near-the-surface electrons.

\begin{figure}[!b]
\centering
\includegraphics[width=3in]{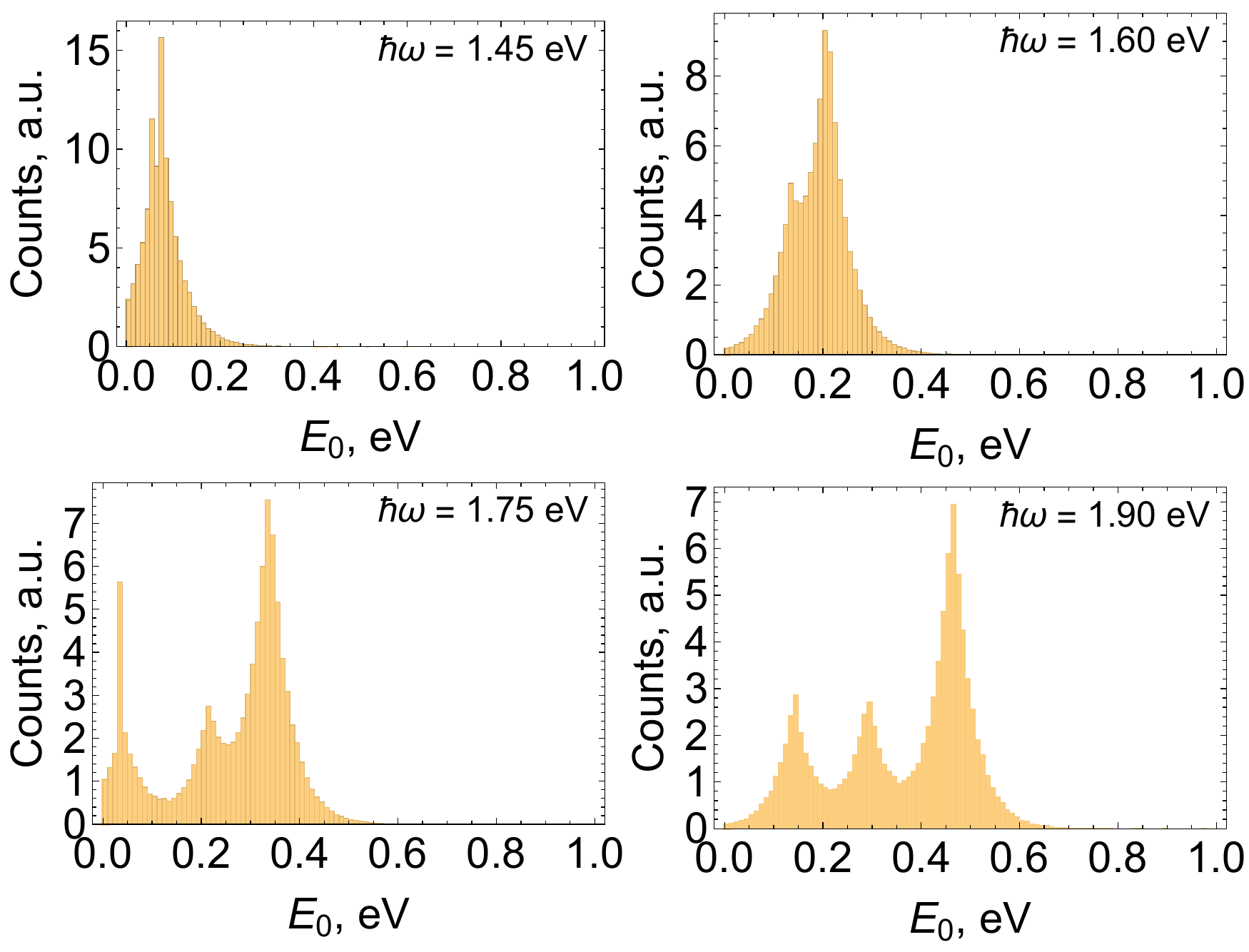}
\caption{Initial energy distribution of electrons photoexcited to the $\Gamma$ valley by photons of different energies. With the photon energy close to the band-gap energy, electrons are photoexcited from the $hh$ and $lh$ sub-bands just above the bottom of the $\Gamma$ valley, where the electrons excited from the $hh$ sub-band have a slightly higher excess energy than those excited from the $lh$ sub-band. When the photon energy becomes large enough ($\hbar\omega > E_\text{g} + \Delta_\text{so}$), transitions from all three sub-bands become possible. Transitions from the $so$ sub-band result in the smallest excess energy. The bin width of the histograms is equal to 10~meV.}
\label{InE}
\end{figure}

The initial excess energy $\Delta E_\text{e}$ of the photoexcited electron can be found from energy conservation $\Delta E_\text{e}+E_\text{g}+\Delta E_\text{h}+\Delta=\hbar \omega$ and momentum conservation $\Delta E_\text{e}(1+\alpha\Delta E_\text{e})m_\text{e}^*=\Delta E_\text{h} m_\text{h}^*$ during the direct transition from the $hh$, $lh$, or $so$ band. In terms of Osman and Ferry's notations,\cite{Osman_1987} it is given by
\begin{equation}
\Delta E_\text{e}=\hbar \omega -E_\text{g}-(\Delta E_\text{h}+\Delta),
\end{equation}
where the hole excess energy $\Delta E_\text{h}$ is given by
\begin{equation}
\begin{split}
\Delta E_\text{h} &= \frac{\Gamma_2}{2\alpha}\Bigg(1-\sqrt{1-\frac{4\alpha(1+\alpha\Gamma_1)\Gamma_1}{\Gamma_2^2}}\Bigg),\\
\Gamma_1 &= \hbar \omega-E_\text{g}-\Delta,\\ 
\Gamma_2 &= 1+m_\text{h}^*/m_\text{e}^*+2\alpha\Gamma_1.
\end{split}
\end{equation}
For an unstrained GaAs, $\Delta$ is zero for transitions from the $hh$ and $lh$ bands and is equal to the spin-off splitting $\Delta_\text{so}$ for the transitions from the $so$ band. The band gap $E_\text{g}$ depends on the doping density and is given by\cite{Zeghbroeck_principles, Tiwari_1990}
\begin{equation}
E_\text{g}=E_{\text{g0}}-\frac{3e^2}{16 \pi\epsilon_\text{s}}\sqrt{\frac{e^2p}{\epsilon_\text{s} k_\text{B} T}}
\label{Eg}
\end{equation}
where $E_{\text{g0}}$ is the intrinsic band gap at room temperature, $e$ is the elementary electronic charge, $\epsilon_\text{s}$ is the static dielectric constant, $p$ is the hole concentration which is simply given by the doping density, $k_\text{B}$ is Boltzmann's constant, and $T$ is the lattice temperature (we assume $T=300$ K for all calculations in this work). Then to account for the warping of the valence band and the width of a laser pulse,\cite{Lugli_1989} the initial energy of the photoexcited electrons can be calculated assuming a small broadening around $\Delta E_\text{e}$ as
\begin{equation}
    E_0 = \Delta E_\text{e} \pm \frac{3}{2} k_\text{B}T\ln(r),
\end{equation}
where the plus or minus sign is chosen randomly for each electron and $r$ is the random number between zero and unity generated for each electron. If the initial energy calculated in this way is negative or equal to zero, we assume that $E_0 = \Delta E_\text{e}$. 
The resulting initial energy distribution of electrons in the $\Gamma$ valley of the CB is shown in Fig.~\ref{InE}.

As it was mentioned in Section~\ref{sec:three-step model}, the initial electron spin polarization $ESP_0$ for the photon energy close to the band-gap energy is equal to 50$\%$ (\ie it is the maximum possible degree of spin polarization determined in the absence of spin relaxation by the selection rules for the transitions at $\bold k = 0$ due to circularly polarized light). With increasing $\hbar\omega$ (transitions away from $\bold k = 0$), in semiconductors with a small spin-orbit splitting $\Delta_\text{so}$ the states of the $so$ sub-band begin to be mixed in with the states of the $lh$ and $hh$ sub-bands.\cite{Dyakonov_1971} Because of this, the relative strength for the photoexcitation of electrons into different spin states of the CB changes and the degree of the initial spin polarization decreases. Dependence of the initial spin polarization $ESP_0$ on the photon energy $\hbar \omega$ can be estimated as\cite{Dyakonov_1971} 
\begin{equation}
ESP_0(\hbar\omega) = \frac{\sum_i P_i K_i}{\sum_iK_i},
\label{eq:P0}
\end{equation}
where the subscript $i=1,2,3$ stands for the $hh$, $lh$, and $so$ sub-band, respectively. Quantities $P_i$ and $K_i$ for a $hh$ sub-band are given by
\begin{equation}
\begin{split}
    P_1 &= \frac{1}{2},\\
    K_1 &= D\sqrt{2x}(3\zeta - 3)^{-3/2},
\end{split}
\end{equation}
and for the $lh$ and $so$ sub-bands
\begin{equation}
\begin{split}
    P_{2,3} &= \frac{(1+\zeta)(\mp g-6x+9\zeta-5)}{4(\pm \zeta g+6x-\zeta-3)},\\
    K_{2,3} &= D\kappa\frac{2x-3\zeta \kappa^2-1}{3\zeta(2x-3\zeta \kappa^2-1)+9\kappa^2-1},
\end{split}
\end{equation}
where $D$ is some constant which is not required to calculate initial polarization $ESP_0$ given by Eq.~\ref{eq:P0}. Parameter $ \kappa$ is given by
\begin{equation}
\begin{split}
    \kappa^2 &= \frac{\pm g+6\zeta x-3\zeta-1}{9(\zeta^2-1)},
\end{split}
\end{equation}
where the upper sign is used for the $lh$ sub-band and the lower sign is used for the $so$ sub-band, and
\begin{equation}
\begin{split}
    g &= \sqrt{36 x^2-12x(\zeta+3)+(3\zeta +1)^2},\\
    x &= \frac{\hbar\omega - E_\text{g}}{\Delta_\text{so}},\\
    \zeta &= \frac{4}{3}\frac{1/m_\text{e}^*+3/(4 m_\text{lh}^*)+1/(4m_\text{hh}^*)}{1/m_\text{lh}^*-1/m_\text{hh}^*}.
\end{split}
\end{equation}

\begin{figure}[!t]
\centering
\includegraphics[width=3in]{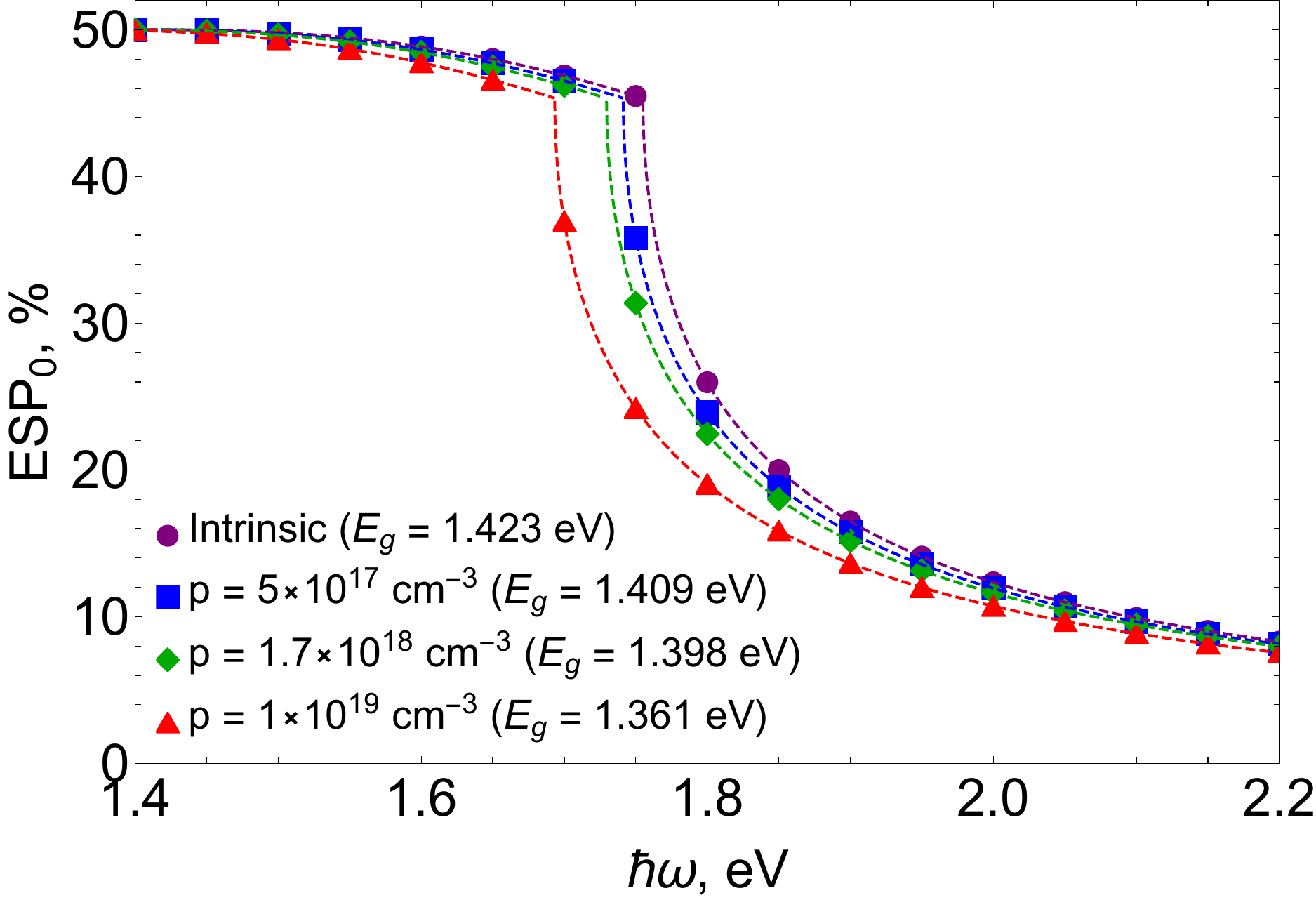}
\caption{Initial, maximum achieved, spin polarization of photoexcited electrons in GaAs as a function of the photoexcitation energy for different doping densities. $ESP_0$ decreases slowly when the transitions away from $\bold{k} = 0$ become possible and then drops fast when the photon energy becomes large enough to photoexcite electrons from the $so$ sub-band.}
\label{P_max}
\end{figure}

The resulting initial spin-polarization curves for several doping densities are shown in Fig.~\ref{P_max} where we assume that the band gap $E_\text{g}$ of GaAs shrinks with the doping density according to Eq.~\ref{Eg}. The initial spin polarization slowly decreases from its peak value of 50$\%$ at the band-gap energy to about 47$\%$ and then drops fast when the optical transitions from the $so$ VB become possible. The initial spin polarization was used to determine the initial spin orientation of photoexcited electrons, \ie the number of electrons photoexcited from the $hh$ sub-band into the spin-up CB state, and from the $lh$ and $so$ sub-bands into the spin-down CB state.

\subsection{\label{sec:transport}Transport}

The electron transport in GaAs is modeled using the Monte Carlo approach\cite{Vasileska_2010} which has been modified to add required scattering processes and to include spin-relaxation mechanisms. Detailed description of the method as well as implementation details can be found elsewhere.\cite{Vasileska_2010,Vasileska_Computational,Tomizawa_Numerical,Jacoboni_1983} Here we only highlight basic concepts.

The Monte Carlo approach for electron transport is a simulation of motion of conductive electrons undergoing numerous scattering events in the crystal under the influence of external forces (\eg an applied electric field $\bold{E}$). The position of each photoexcited electron is tracked in both real and $k$-space using the semi-classical approach
\begin{equation}
\begin{split}
\frac{d\bold{r}}{dt} &= \bold{v},\\
\end{split}
\end{equation}
\begin{equation}
\begin{split}
\hbar\frac{d\bold{k}}{dt} &= -e \bold{E},
\end{split}
\end{equation}
where the wave vector $\bold{k}$ is related to the group velocity $\bold{v}$ through
\begin{equation}
\bold{v} = \frac{\hbar\bold k}{m_\text{e}^*} \Big( \frac{d\gamma_{\bold k}}{dE_{\bold k}}\Big)^{-1} =\frac{\hbar\bold k}{m_\text{e}^*}\frac{1}{(1+2\alpha E_{\bold k})}.
\label{velocity}
\end{equation} 

Electrons interact with the lattice, defects, and other carriers through different scattering mechanisms which may change the energy and momentum of the particle and cause a spin flip. The probability of each scattering mechanism is given by its scattering rate and the probability of a spin-flip in that scattering is defined by the spin relaxation rate. The Monte Carlo technique includes generation of a free flight time, choice of the type of scattering event that occurs at the end of a free flight, and recalculation of the energy and momentum as well as spin orientation after the scattering for each electron. The procedure then repeats for the next free flight and stops after the total time exceeds the simulation time.  

\subsubsection{\label{sec:momentum_relaxation_mechanisms}Momentum Relaxation Mechanisms}

In general, the transition rate (or the transition probability per unit time from the initial state $\bold k$ to the final state $\bold k^{'}$) for a scattering process is given by Fermi's Golden rule
\begin{equation}
S(\bold k,\bold k^{'}) = \frac{2 \pi}{\hbar} \Big|\langle\bold k^{'}|H^{'}|\bold k\rangle\Big|^2\delta(E_{\bold k^{'}}-E_{\bold k}\mp \hbar\omega),
\label{golgen_rule}
\end{equation}
where the $\delta$-function expresses the conservation of energy and $\hbar\omega$ is the absorbed (upper sign) or emitted (lower sign) energy. Then the scattering rate $W(\bold k)$ and the momentum relaxation rate $1/\tau_\text{m}(\bold k)$ can be derived as
\begin{equation}
\begin{split}
W(\bold k) &= \frac{\Omega}{(2\pi)^3}\int S(\bold k,\bold k^{'}) d\bold k^{'},
\end{split}
\end{equation}

\begin{equation}
\frac{1}{\tau_\text{m}(\bold k)} = \frac{\Omega}{(2\pi)^3}\int S(\bold k,\bold k^{'})\Big(1-\frac{\bold k^{'}}{\bold k}\cos\theta\Big) d\bold k^{'},
\end{equation}
where $\Omega$ is the volume of the crystal and $\theta$ is the angle between initial $\bold{k}$ and final $\bold{k}'$ states. $\tau_\text{m}(\bold k)$ is known as the momentum relaxation time and is the time required to randomize the momentum.\cite{Lundstrom_Fundamentals} In the simulation, we consider phonon scattering (scattering by acoustic and optical phonons), defect scattering (scattering by ionized impurities), and binary electron-hole interactions. Below, we provide scattering rates and momentum relaxation rates for the mechanisms implemented in our model. 

\begin{itemize}
\item \textbf{Acoustic phonon scattering.}
Acoustic phonons result from coherent oscillations of neighboring atoms of the lattice. The total intravalley scattering rate that accounts for absorption and emission of an acoustic phonon in a non-parabolic band approximation is given by\cite{Ridley_Quantum}
\begin{equation}
W_\text{ap}(\bold{k})=\frac{\sqrt{2}(m_{\text{e}}^*)^{3/2}\Xi_\text{d}^2k_\text{B}T
\sqrt{\gamma_{\bold k}}}{c_\text{l}\pi\hbar^4}(1+2\alpha E_{\bold k}),
\label{ap_sc_rate}
\end{equation}
where $\Xi_\text{d}$ is the deformation potential. $c_\text{l}$ is the material elastic constant given by $c_\text{l}=\rho v_\text{s}^2$, where $\rho$ is the crystal density and $v_\text{s}$ is the sound velocity. $\gamma_{\bold k}$ is given by Eq.~\ref{disp_CB}. This is an isotropic process, \ie it has a uniform probability of scattering in all directions. Therefore, the final wave vector after scattering is defined by the scattering angle $\theta$ calculated using a random number $r$ between zero and one
\begin{equation}
    \cos\theta = 1 - 2 r.
\end{equation}
For the case of isotropic acoustic phonon scattering, the momentum relaxation rate $1/\tau_\text{m}^\text{ap}(\bold{k})$ is given by the scattering rate $W_\text{ap}(\bold{k})$, Eq.~\ref{ap_sc_rate}.
 
\item \textbf{Screened polar optical phonon scattering.}
The oscillations of neighboring atoms in a lattice in the opposite directions give rise to polar optical phonon scattering. The scattering rate for absorption ($W_\text{pop}(\bold k)\propto N_0$) and emission ($W_\text{pop}(\bold k)\propto N_0+1$) of a polar phonon of energy $\hbar \omega_0$ is given by\cite{Ridley_Quantum}
\begin{equation}
\begin{split}
W_\text{pop}(\bold k) &= \frac{e^2(\hbar \omega_0)}{8\pi\hbar^2\epsilon_\text{p}}\frac{\sqrt{m_\text{e}^*}(1+2\alpha E_\bold{k'})}{\sqrt{2\gamma(E_\bold{k})}} \big(N_\text{0}+\frac{1}{2}\mp\frac{1}{2} \big)\\
&\quad\times
\Bigg[\Big(\frac{\beta^2}{\beta^2+q_\text{max}^2(E_\bold{k})}-\frac{\beta^2}{\beta^2+q_\text{min}^2(E_\bold{k})}\Big)\\
&\quad +\ln\Big|\frac{\beta^2+q_\text{max}^2(E_\bold{k})}{\beta^2+q_\text{min}^2(E_\bold{k})}\Big|\Bigg],
\end{split}
\end{equation}
where $\epsilon_\text{p} = (1/\epsilon_{\infty}-1/\epsilon_\text{s})^{-1}$ is the effective dielectric constant that combines the high-frequency dielectric constant $\epsilon_{\infty}$ and the static dielectric constant $\epsilon_\text{s}$, $\beta = 1/L$ is the inverse screening length, and $N_0$ is the phonon distribution function which at the thermodynamic equilibrium is simply given by the Bose-Einstein function 
\begin{equation}
N_{0}=\frac{1}{\exp\big[\hbar \omega_0/(k_\text{B} T)\big]-1}.
\end{equation}
The minimum and maximum values of a phonon wavevector $q$ are given by
\begin{equation}
\begin{split}
q_\text{min}(E_\bold{k})&=\frac{\sqrt{2m_\text{e}^*\gamma(E_\bold{k})}}{\hbar}\Bigg(\pm\sqrt{\frac{\gamma(E_\bold{k'})}{\gamma(E_\bold{k})}}\mp 1\Bigg),\\
q_\text{max}(E_\bold{k})&=\frac{\sqrt{2m_\text{e}^*\gamma(E_\bold{k})}}{\hbar}\Bigg(\sqrt{\frac{\gamma(E_\bold{k'})}{\gamma(E_\bold{k})}}+ 1\Bigg).
\end{split}
\end{equation}
Here $E_\bold{k'}=E_\bold{k}\pm\hbar\omega_0$, and the upper and lower signs stand for the absorption and emission of a phonon, respectively. 
For the nondegenerate case, the screening length $L$ is given by the Debye length
\begin{equation}
L_\text{D}=\sqrt{\frac{\epsilon_\text{s} k_\text{B} T}{e^2 p}}, 
\label{debye}
\end{equation}
For degenerate semiconductors, the approach which assumes the Thomas-Fermi screening by mobile light holes is more appropriate\cite{Taniyama_1990} 
\begin{equation}
\L_{\text{TF}} = \sqrt{\frac{\pi \hbar^2 \epsilon_\text{s}}{e^2 m_\text{lh}^*}}\Big(\frac{\pi}{3 N_\text{lh}}\Big)^{1/6},
\label{thomas_fermi}
\end{equation}
where the concentration of light holes can be estimated as $N_\text{lh} = m_\text{lh}^{*3/2} / (m_\text{hh}^{*3/2} + m_\text{lh}^{*3/2}) p$.

For this anisotropic mechanism, the angle between final and initial states can be calculated using a random number $r$ \cite{Vasileska_2010}
\begin{equation}
\begin{split}
\cos\theta &= \frac{(1+\xi)-(1+2\xi)^r}{\xi},\\
\xi &= \frac{2\sqrt{\gamma_{\bold k}\gamma_{\bold k'}}}{(\sqrt{\gamma_{\bold k}}-\sqrt{\gamma_{\bold k'}})^2}.
\end{split} 
\end{equation}
The momentum relaxation rate can be calculated as \cite{Ridley_Quantum}
\begin{equation}
\begin{split}
\frac{1}{\tau_m^{\text{pop}}(\bold k) }
&= \frac{e^2(\hbar \omega_0)}{16\pi\hbar^2\epsilon_\text{p}}\frac{\sqrt{m_\text{e}^*}(1+2\alpha E_\bold{k'})}{\sqrt{2\gamma(E_\bold{k})}} \big(N_\text{0}+\frac{1}{2}\mp\frac{1}{2} \big)\\
&\quad\times \Bigg[\pm\frac{N_1}{E_\bold{k} (q_\text{min}^2+\beta^2)} \mp \frac{N_2}{E_\bold{k} (q_\text{max}^2+\beta^2)}\Bigg],\\
N_1 &= \hbar \omega_0 \beta^2\mp \frac{E_\bold{k}E_{\beta}(q_\text{min}^4/\beta^2+2q_\text{min}^2)}{\gamma(E_\bold{k})}\\
&\quad +(q_\text{min}^2+\beta^2)\Big(\hbar \omega_0\pm \frac{2E_\bold{k}E_{\beta}}{\gamma(E_\bold{k})}\Big)\ln\Big|q_\text{min}^2+\beta^2\Big|,\\
N_2 &= \hbar \omega_0 \beta^2\mp \frac{E_\bold{k}E_{\beta}(q_\text{max}^4/\beta^2+2q_\text{max}^2)}{\gamma(E_\bold{k})}\\
&\quad +(q_\text{max}^2+\beta^2)\Big(\hbar \omega_0 \pm \frac{2E_\bold{k}E_{\beta}}{\gamma(E_\bold{k})}\Big)\ln\Big|q_\text{max}^2+\beta^2\Big|.
\label{pop_rel_rate}
\end{split}
\end{equation}
where
\begin{equation}
E_{\beta} = \frac{\hbar^2\beta^2}{2m_\text{e}^*}.
\end{equation}

\item \textbf{Intervalley scattering.}
Intervalley transitions from valley $i$ to valley $j$ are described by the absorption ($W_{ij}(\bold k)\propto N$) or emission ($W_{ij}(\bold k)\propto N+1$) of non-polar optical phonons\cite{Vasileska_Computational}
\begin{equation}
\begin{split}
W_{ij}(\bold k) &= \frac{(m_{\text{e}_j}^*)^{3/2}D_{ij}^2 Z_j\sqrt{\gamma_{\bold k'}}}{\sqrt{2}\pi\rho \hbar^2(\hbar \omega_{ij})(1+2\alpha_j E_{\bold k'})}\big(
N+\frac{1}{2}\mp\frac{1}{2}\big),
\end{split}
\label{i_sc_rate}
\end{equation}
where $D_{ij}$ is the intervalley deformation potential, $Z_j$ is the number of equivalent final valleys for the electron to scatter into, and $N$ is the Bose-Einstein distribution function for the phonons of energy $\hbar \omega_{ij}$ involved in the scattering process
\begin{equation}
N = \frac{1}{\exp\big[\hbar \omega_{ij}/(k_\text{B} T)\big]-1}.
\end{equation}
Final energy after absorption (upper sign) or emission (lower sign) of a phonon is given by $E_{\bold k'}=E_{\bold k}\pm\hbar \omega_{ij}-\Delta_{ji}$, where $\Delta_{ji}$ is the potential energy difference between the bottoms of valleys $j$ and $i$. This is an isotropic process so the momentum relaxation rate $1/\tau_\text{m}^{ij}(\bold k)$ is given by the scattering rate $W_{ij}(\bold k)$, Eq.~\ref{i_sc_rate}.

\item \textbf{Ionized impurity scattering.}
In $p$-doped semiconductors, minor electrons scatter from negatively-charged ions via Coulomb-like potential screened by mobile majority carriers (holes). In the Brooks-Herring approach, the scattering rate for this elastic mechanism is given by \cite{Vasileska_Computational}
\begin{equation}
\begin{split}
W_{\text{ii}}(\bold k) = \frac{N_\text{a}^- e^4Z^2(2m_\text{e}^*)^{3/2}}{2\pi\hbar^4 \epsilon_{\text{s}}^2 \beta^4} \frac{\sqrt{\gamma_{\bold k}}(1+2\alpha E_{\bold k})}{1+4\gamma_{\bold k}/E_{\beta}},
\end{split}
\label{imp_sc_rate}
\end{equation}
where $N_\text{a}^-$ is the 
density of negatively charged acceptor ions, and the product $eZ$ is the ionized impurity charge. We assume full ionization. The final angle for scattering with ionized impurities is given by \cite{Vasileska_Computational}
\begin{equation}
\cos\theta = 1 - \frac{2 r}{1+4 \gamma_{\bold k} (1-r)/E_{\beta}},
\label{imp_sc_angle}
\end{equation}
where $r$ is a random number between zero and one. The corresponding momentum relaxation rate can be calculated as \cite{Tomizawa_Numerical}
\begin{equation}
\begin{split}
\frac{1}{\tau_\text{m}^{\text{ii}}(\bold k)} &= \frac{N_\text{a}^- e^4Z^2}{16\pi\epsilon_\text{s}^2\sqrt{2m_\text{e}^*}}\frac{(1+2\alpha E_{\bold k})}{\gamma_{\bold k}^{3/2}}\\
&\quad\times\Big(\ln\big|1+\frac{4\gamma_{\bold k}}{E_{\beta}}\big| - \frac{4\gamma_{\bold k}/E_{\beta}}{1+4\gamma_{\bold k}/E_{\beta}} \Big).
\end{split}
\label{imp_rel_rate}
\end{equation}

\item \textbf{Electron-hole scattering.}
Electron-hole scatterings also play a significant role in a minority electron transport in $p$-doped GaAs and provide an additional channel for energy transfer. This interaction is implemented using the rejection technique\cite{Brunetti_1985} where the scattering rate is replaced by its maximum value given by the constant
\begin{equation}
    W_\text{eh}^\text{max}(\bold k) =\frac{p m_\text{R} e^4}{4\pi \epsilon_\text{s}^2\hbar^3\beta^3},
\end{equation}
where 
\begin{equation}
    m_\text{R} = \frac{m_\text{e}^*m_\text{h}^*}{m_\text{e}^* + m_\text{h}^*}
\end{equation}
is the reduced mass of an electron and a hole. When an electron-hole scattering is chosen within the simulation, an interaction with a randomly chosen hole is accepted only if a random number $r$ between 0 and 1 satisfies the following inequality
\begin{equation}
    r < \frac{2 g \beta}{g^2+\beta^2},
\end{equation}
where
\begin{equation}
    g =2m_\text{R}|\bold{k}_0/m_\text{h}^*-\bold{k}/m_\text{e}^*|
\end{equation}
and $\bold{k}_0$ and $\bold{k}$ are the wave vectors of colliding hole and electron, respectively. If we define $\bold{g} = 2 m_\text{R}(\bold{k}_0/m_\text{h}^*-\bold{k}/m_\text{e}^*)$ as a wave vector difference of electron and hole before the scattering and $\bold{g'} = 2m_\text{R}(\bold{k'}_0/m_\text{h}^*-\bold{k'}/m_\text{e}^*)$ as a wave vector difference of electron and hole after the scattering, the angle $\vartheta$ between vectors $\bold{g}$ and $\bold{g'}$ is given by
\begin{equation}
    \cos\vartheta = 1 - \frac{2 r}{1+g^2 (1-r)/\beta^2}.
\end{equation}
Using this expression and assuming that the azimuthal angle is randomly distributed between 0 and $2\pi$, and $g = |\bold{g}| = |\bold{g'}|$, the relative wave vector $\bold{g'}$ after the scattering can be obtained using a standard technique.\cite{Tomizawa_Numerical,Vasileska_Computational} Then the wave vector $\bold{k'}$ of the scattered electron can be calculated as
\begin{equation}
    \bold{k'} = \bold{k} - (\bold{g'} - \bold{g})/2.
\end{equation}

\item \textbf{Pauli exclusion principle.} In heavily $p$-doped semiconductors, the degeneracy of holes has a significant effect on the transport properties of electrons because the Fermi level in this case lies in the VB and many hole states are occupied.\cite{Lugli_1985,Furuta_1990,Taniyama_1990} To implement the Pauli exclusion principle, we follow a simple approach.\cite{Taniyama_1990} If a final state of the hole, participating in an electron-hole scattering, has a lower energy than the Fermi energy given by
\begin{equation}
    E_\text{F}^h = \frac{(3\pi^2\hbar^3p)^{2/3}}{2m_\text{h}^*},
\end{equation}
the scattering event is rejected and treated as a self-scattering. Degeneracy effect reduces the electron-hole scattering due to Pauli exclusion principle.

\end{itemize}

\begin{figure}[!t]
\centering
\subfigure[][]{\includegraphics[width=3in]{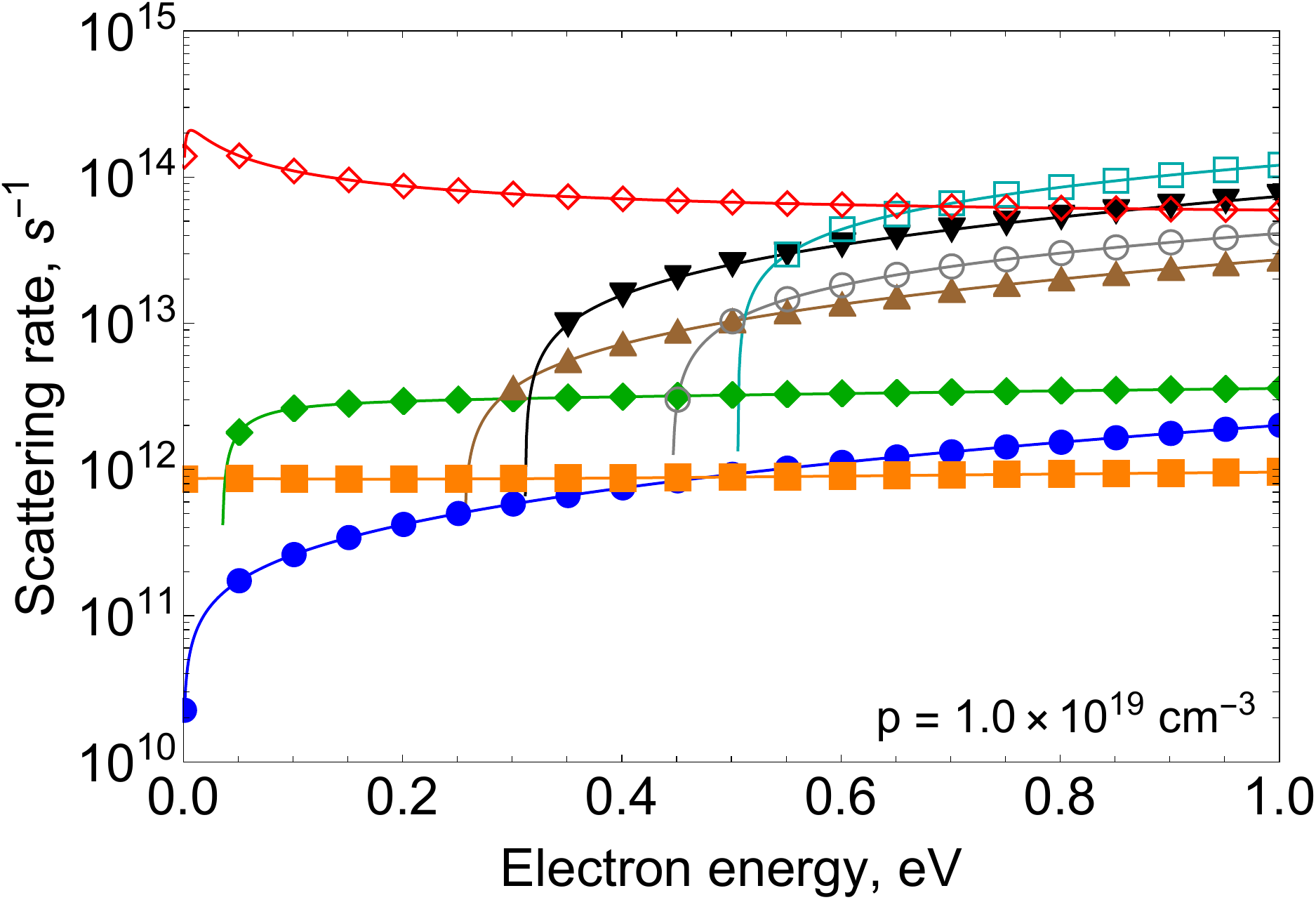}
\label{gamma_15}}
\hfil
\subfigure[][]{\includegraphics[width=3in]{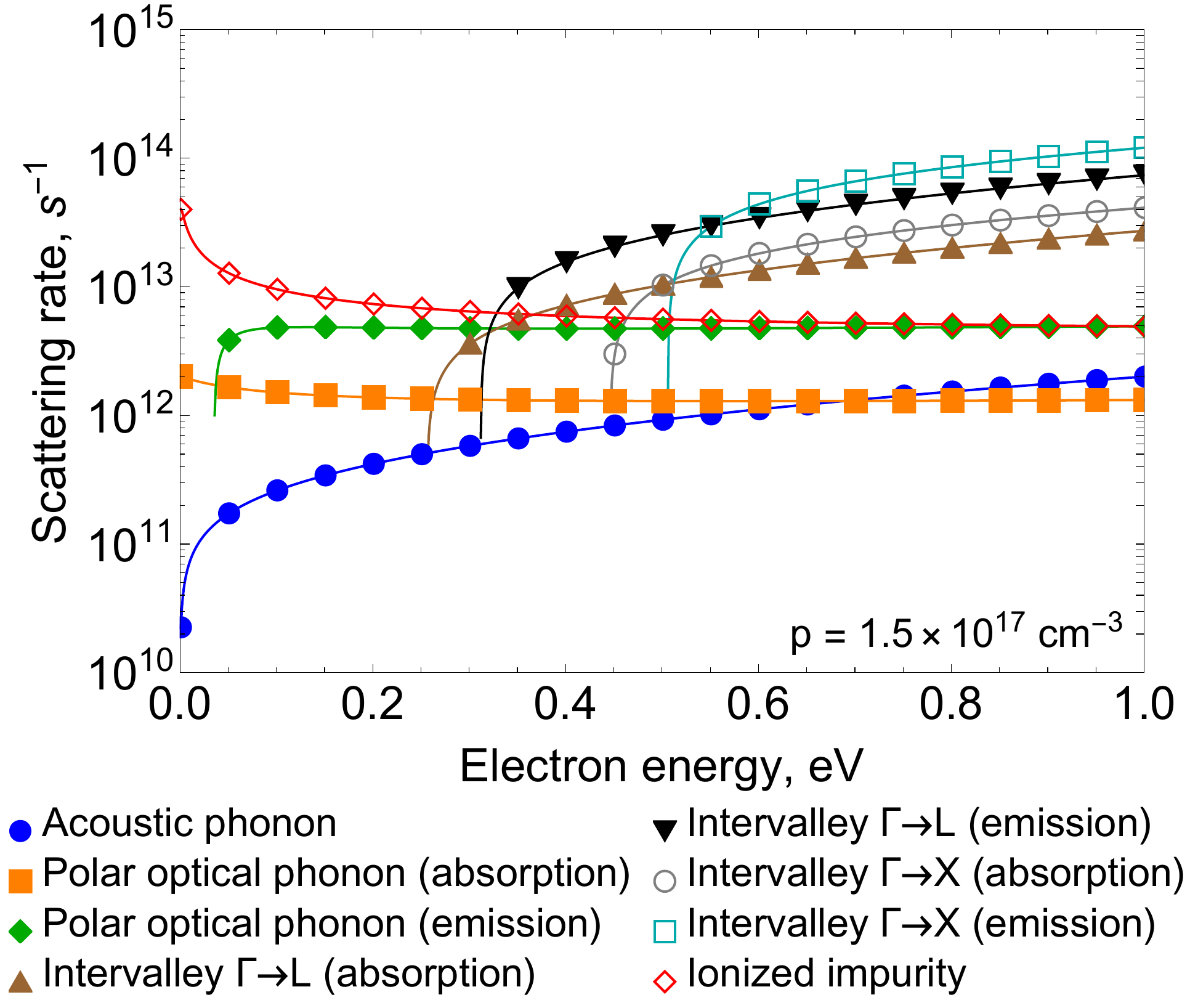}
\label{gamma_19}}
\caption{Scattering rates for electrons in the $\Gamma$ valley of (a) heavily doped ($p = 10^{19}$ cm$^{-3}$) and (b) moderately doped ($p =1.5 \times 10^{17}$ cm$^{-3}$) GaAs calculated as a function of the electron energy measured with respect to the bottom of the valley.}
\label{gamma_17_19}
\end{figure}

\begin{figure}[!t]
\centering
\subfigure[][]{\includegraphics[width=3in]{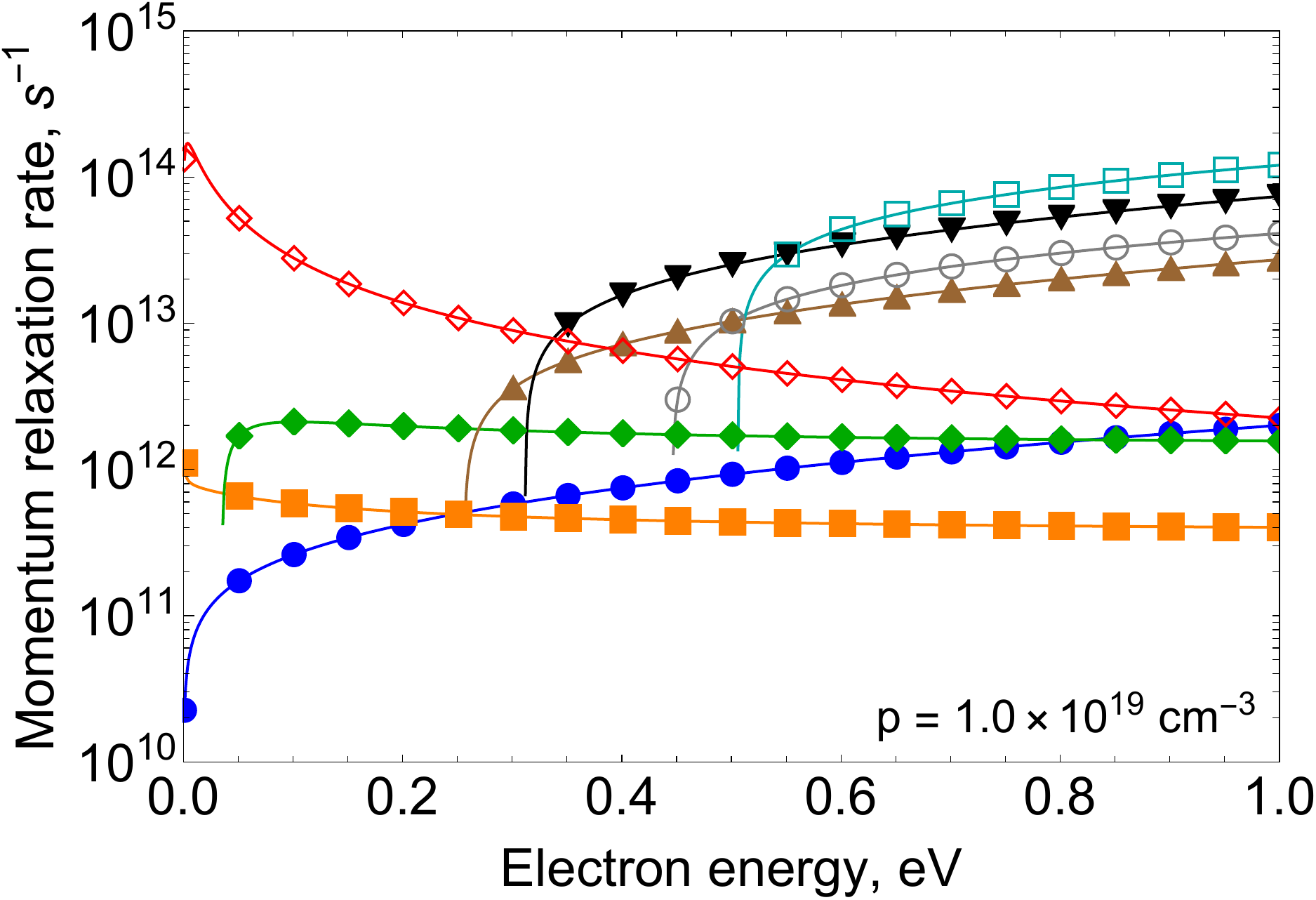}
\label{L_19}}
\hfil
\subfigure[][]{\includegraphics[width=3in]{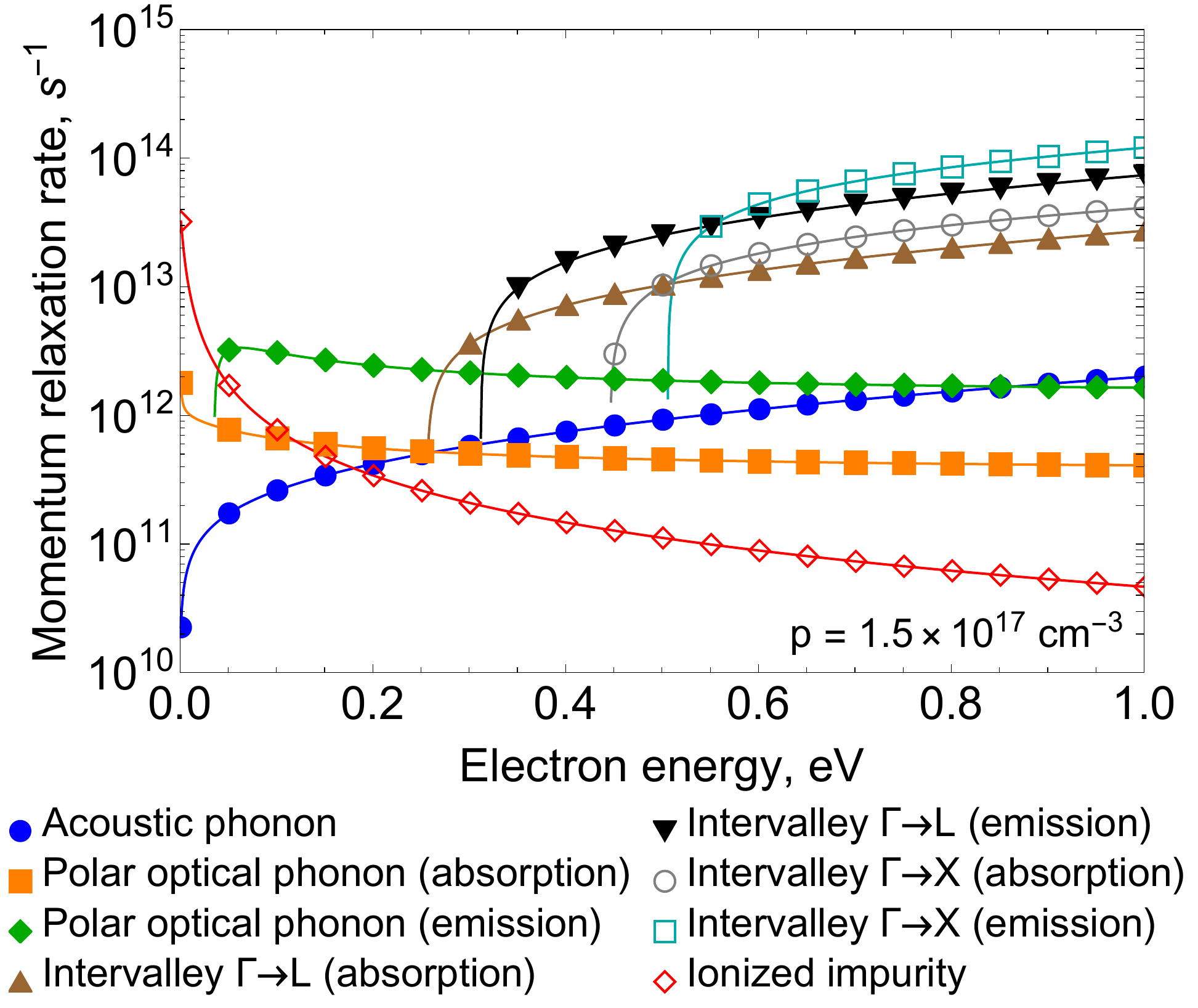}
\label{X_19}}
\caption{Momentum relaxation rates for electrons in the $\Gamma$ valley of (a) heavily doped ($p = 10^{19}$ cm$^{-3}$) and (b) moderately doped ($p = 1.5\times 10^{17}$ cm$^{-3}$) GaAs calculated as a function of the electron energy measured with respect to the bottom of the valley.}
\label{gamma_m_relax_17_19}
\end{figure}

Calculated scattering rates and momentum relaxation rates for the $\Gamma$ valley are shown in Figs.~\ref{gamma_17_19} and \ref{gamma_m_relax_17_19}, respectively, for two doping densities. In GaAs the hole mass exceeds the electron mass significantly, so for simplicity we avoid simulation of the hole transport and consider holes to be at rest. We consider only the interactions with heavy holes since they have a higher density of states as compared to light holes.\cite{Taniyama_1990} We also assume that the hole concentration is given by the concentration of acceptor atoms, therefore $p=N_\text{a}^-=N_\text{a}$. 

To check the validity of our approximations and chosen material parameters, we model electron drift velocity under influence of the applied electric field. For the purpose of drift-velocity calculations, we assume thermalized Maxwellian energy distributions for electrons, whereas in the full Monte Carlo simulation of photoemission the initial energy distribution of photoexcited electrons depends on the photon energy as shown in Fig.~\ref{InE}.

\begin{figure}[!t]
\centering
\includegraphics[width=3.in]{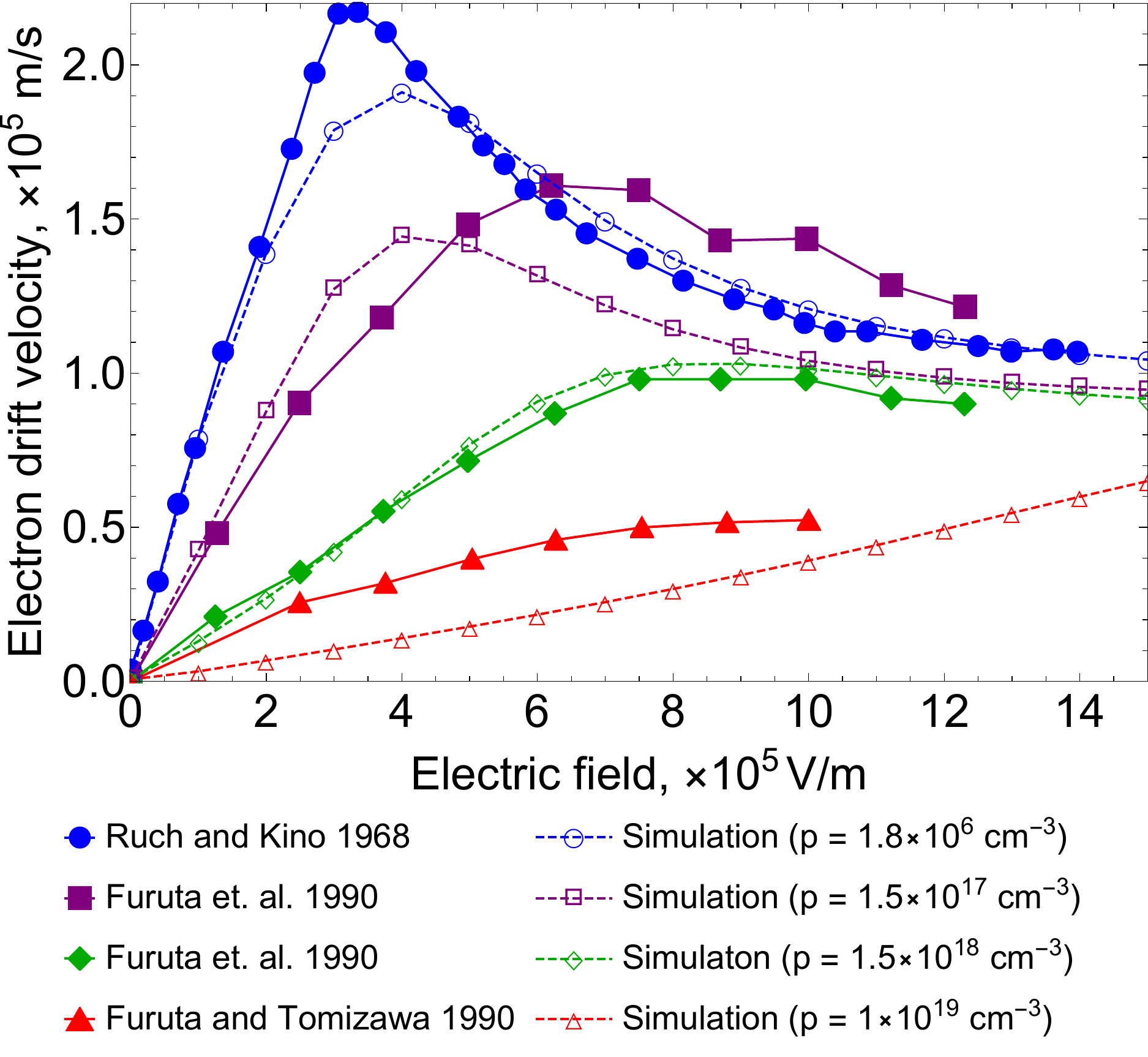}
\caption{Electron drift velocity as a function of the applied electric field in a $p$-doped GaAs compared to the experimental data\cite{Ruch_1968, Taniyama_1990} for various doping densities.}
\label{drift}
\end{figure}

Our drift velocity results for various doping densities are compared to the experimental data\cite{Ruch_1968, Taniyama_1990} in Fig.~\ref{drift}. At small electron energies in heavily doped samples, ionized impurity and hole scatterings are dominant mechanisms. Both the scattering rate and the scattering angle increase with increasing doping density. Therefore, the electron mobility, the proportionality coefficient between the drift velocity and the applied electric field
, tends to be smaller for heavily doped materials. This effect can be seen in Fig. \ref{drift}, where the slope of the drift-velocity curves in the low-energy region decreases with increasing doping density. As the electron energy increases, both impurity scattering rate and scattering angle slightly decrease, and other scattering mechanisms start playing significant roles. Electrons lose energy through inelastic interactions with polar optical phonons and intervalley scatterings. This is so-called saturation region on the drift-velocity plot.

\subsubsection{\label{sec:spin_relaxation_mechanisms}Spin Relaxation Mechanisms}

Depolarization, or relaxation of electron spin states, occurs via the interaction with other carriers, impurities, acoustic and polar optical phonons. We consider three main mechanisms that lead to the spin relaxation of spin-polarized electrons in the CB of GaAs: the Elliott-Yafet (EY)\cite{Elliott_1954,Yafet_1963}, D'yakonov-Perel (DP)\cite{D'yakonov_1972}, and Bir-Aronov-Pikus (BAP)\cite{Bir_1976} processes. The first two mechanisms are related to the features of the CB structure and the last one considers the exchange interactions between electrons and holes. Below, we briefly discuss these processes and provide the spin relaxation rates used in the calculations. 

\begin{itemize}

\item \textbf{Elliott-Yafet mechanism.}
As described above, the CB S$_{1/2}$ state is two-fold degenerate due to the spin-orbit interaction and the resulting electron eigenfunction is represented as a superposition of spin-up and spin-down components. Consequently, the matrix elements describing scattering mechanisms in the CB involve the terms associated with both intrastate (spin-conserving) and interstate (spin-flip) transitions.\cite{Ridley_Quantum} At the point $\textbf{k}\neq 0$, the probability of spin-flip transitions arises even if the involved scattering mechanism is spin-independent. The Elliott-Yafet mechanism takes into account the mixing of wave functions with different spins as a result of spin-orbit interaction. The corresponding spin relaxation rate $1/\tau_\text{s}^\text{EY}(\bold k)$ is proportional to the momentum relaxation rate $1/\tau_\text{m}^{i}(\bold k)$ and is given by \cite{Fishman_1977}

\begin{equation}
\begin{split}
\frac{1}{\tau_\text{s}^{\text{EY}}(\bold k)} &= A_i \Bigg( 1-\frac{m_\text{e}^*}{m_0}\Bigg)^2 \Bigg(\frac{\eta}{1+\eta}\Bigg)^2 \Bigg(\frac{1+\eta/2}{1+2\eta/3}\Bigg)^2 \\ &\quad\times\Bigg(\frac{E_{\bold k}}{E_\text{g}}\Bigg)^2\frac{1}{\tau_\text{m}^i(\bold k)},
\end{split}
\label{EYeq}
\end{equation}
where $\eta=\Delta_{\text{so}}/E_\text{g}$. The dimensionless coefficient $A_i$ depends on the scattering mechanism $i$. We assume that $A = 32/27$ for all mechanisms.\cite{Fishman_1977}

\item \textbf{D'yakonov-Perel mechanism.}
The D'yakonov-Perel spin relaxation mechanism arises due to the lack of an inversion center in some semiconductors which leads to splitting of the spin states of the CB at $\bold{k}\neq 0$. The corresponding spin relaxation rate is given by\cite{Fishman_1977,Dyson_2004}

\begin{equation}
\begin{split}
\frac{1}{\tau_\text{s}^{\text{DP}}(\bold k)} &= \frac{128}{945} Q_i \frac{\Delta_{\text{so}}^2 B^2}{(1+\eta)(1+2\eta/3)}\frac{{m_\text{e}^*}^2}{\hbar^6}\Bigg(1-\frac{m_\text{e}^*}{m_0}\Bigg)\\
&\quad\times \Bigg(\frac{E_{\bold k}}{E_\text{g}}\Bigg)^3\tau_\text{m}^i(\bold k),
\end{split}
\label{DPeq}
\end{equation}
where the dimensionless factor $Q_i$ varies depending on a scattering mechanism $i$. $Q=1$ for isotropic processes and $Q=1/6$ for anisotropic processes.\cite{Fishman_1977} Since the dominant scattering mechanisms are anisitropic, we use $Q=1/6$ for all scatterings. Parameter $B$ for GaAs is given by $B=10 \hbar^2 / (2 m_0)$.

\item \textbf{Bir-Aronov-Pikus mechanism.}
As it was already pointed out, the BAP mechanism originates from the exchange interaction of electrons with holes. In the case of a non-degenerate semiconductor and assuming that all holes are free, the BAP spin relaxation rate is given by\cite{Pikus_1984, Aronov_1983}
\begin{equation} 
\frac{1}{\tau_\text{s}^{\text{BAP}}(\bold k)} = \frac{2}{\tau_0}\frac{v_{\bold k}}{v_\text{B}} a_\text{B}^3 p |\psi(0)|^4,
\end{equation}
where $v_{\bold k}$ is the magnitude of electron velocity given by Eq.~\ref{velocity}, $v_\text{B}=\hbar/(m_\text{R} a_\text{B})$ is the exciton Bohr velocity, $a_\text{B}=4\pi\hbar^2 \epsilon_\text{s}/(q^2 m_\text{R})$ is the exciton Bohr radius, and $p$ is the concentration of free holes. $\tau_0$ is given by
\begin{equation}
\frac{1}{\tau_0}=\frac{3}{64}\frac{\pi \Delta_{\text{exc}}^2}{\hbar E_\text{B}},
\end{equation}
where $\Delta_{\text{exc}}$ is the exchange splitting of the exciton ground state and $E_\text{B}=\hbar^2/(2 m_\text{R} a_\text{B}^2)$ is the Bohr energy. The Sommerfeld factor for an unscreened Coulomb potential is given by
\begin{equation}
\big|\psi(0)\big|^2=\frac{2\pi}{\varkappa}\Big[1-\exp(-\frac{2\pi}{\varkappa})\Big]^{-1},
\end{equation}
where $\varkappa=\sqrt{E_{\bold k}/E_\text{B}}$. However, in the case of screening it has some more complicated form.\cite{Zerrouati_1988,Bir_1976} In this work, we use $\big|\psi(0)\big|^2 =1 $, which corresponds to the total screening.\cite{Jiang_2009} We also assume that the main contribution to the BAP spin relaxation rate comes from the interactions of electrons with heavy holes since their concentration exceeds the concentrations of holes in the $lh$ and $so$ subbands.\cite{Pikus_1984} 
The BAP spin relaxation rate for a degenerate semiconductor in the case of thermalized electrons ($E_\bold{k} \leq m_\text{h}^* (k_\text{B}T)^2/(m_\text{e}^*E_\text{F}^h)$) is given by\cite{Pikus_1984, Aronov_1983}
\begin{equation} 
\frac{1}{\tau_\text{s}^{\text{BAP}}(\bold k)} = \frac{3}{\tau_0}\frac{v_{\bold k}}{v_\text{B}}\frac{k_\text{B}T}{E_\text{F}^h}a_\text{B}^3 p |\psi(0)|^4.
\end{equation}
For hot electrons with $E_\bold{k} > m_\text{h}^* (k_\text{B}T)^2/(m_\text{e}^*E_\text{F}^h)$, the following expression is used instead
\begin{equation} 
\frac{1}{\tau_\text{s}^{\text{BAP}}(\bold k)} = \frac{2}{\tau_0}\frac{v_\text{F}}{v_\text{B}}\frac{E_\bold{k}}{E_\text{F}^h}a_\text{B}^3 p |\psi(0)|^4,
\end{equation}
where $v_\text{F} = \sqrt{2 E_\text{F}^h/m_\text{h}^*}$ is the Fermi velocity of the holes.

\end{itemize}

\begin{figure}[!b]
\centering
\subfigure[][]{\includegraphics[width=3in]{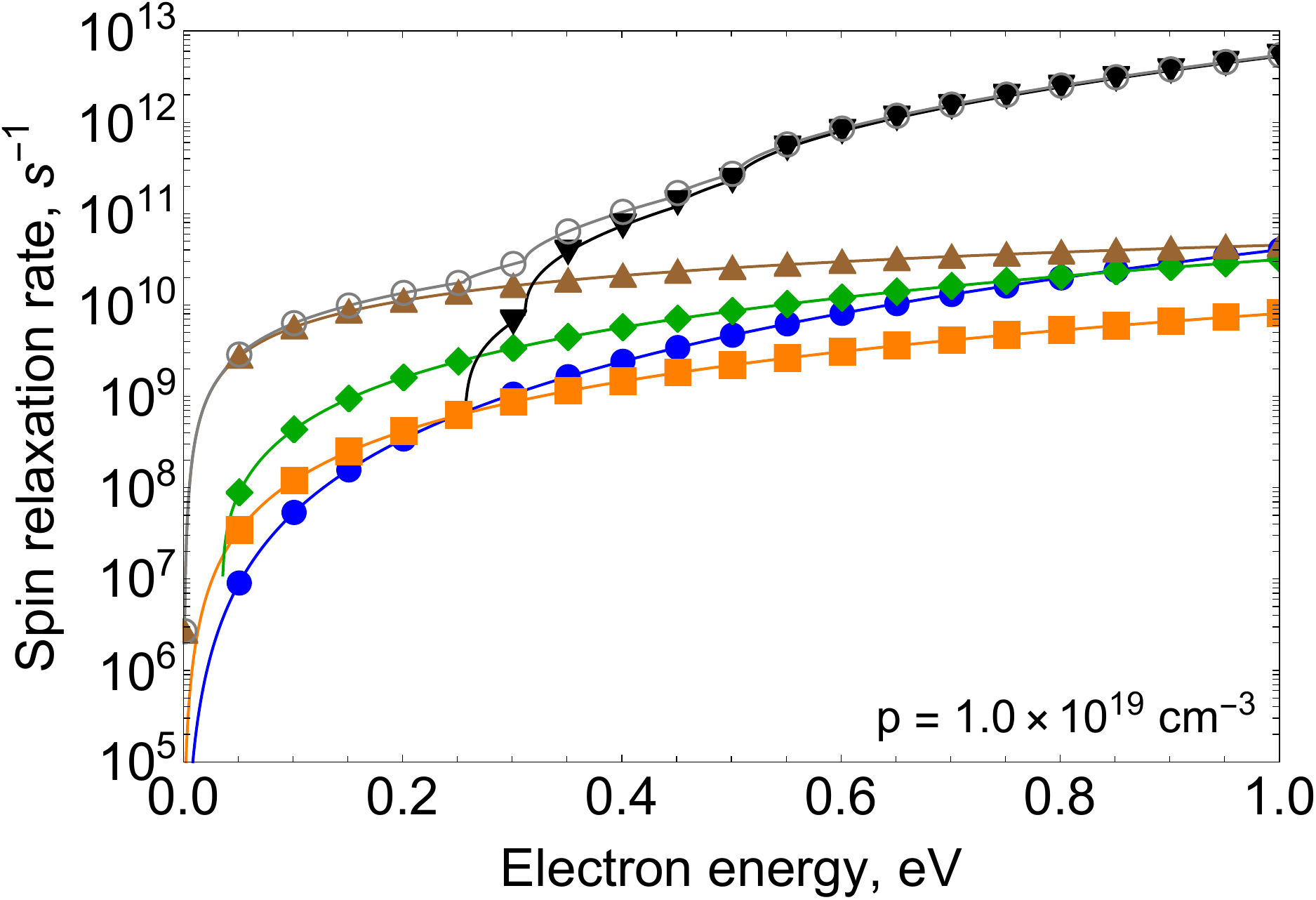}
\label{gamma_EY_19}}
\subfigure[][]{\includegraphics[width=3in]{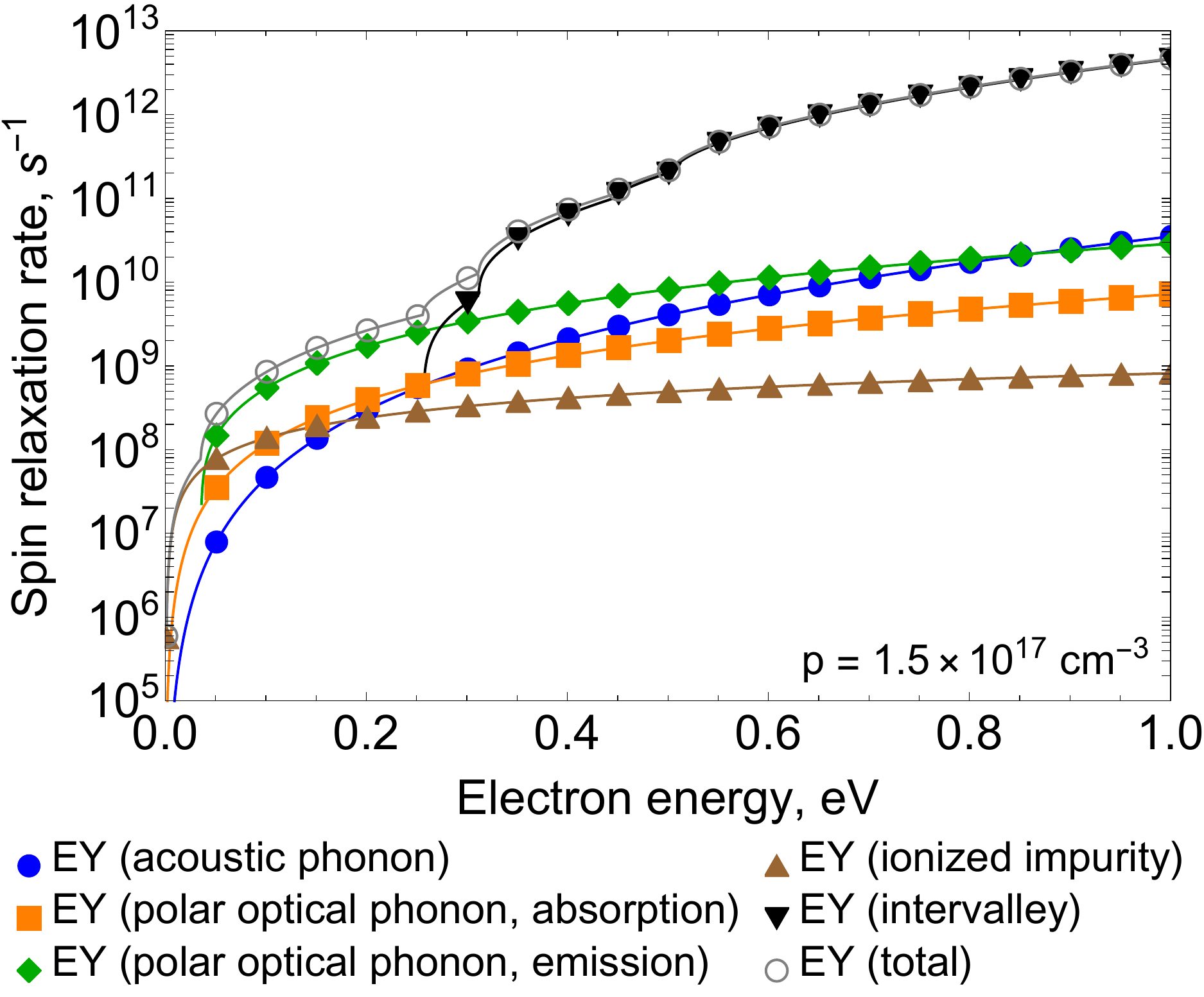}
\label{gamma_EY_17}}
\caption{Elliott-Yafet spin relaxation rates due to different scattering mechanisms in the $\Gamma$ valley of (a) heavily doped and (b) moderately doped GaAs.}
\label{gamma_EY_19_17}
\end{figure}

\begin{figure}[!t]
\centering
\subfigure[][]{\includegraphics[width=3in]{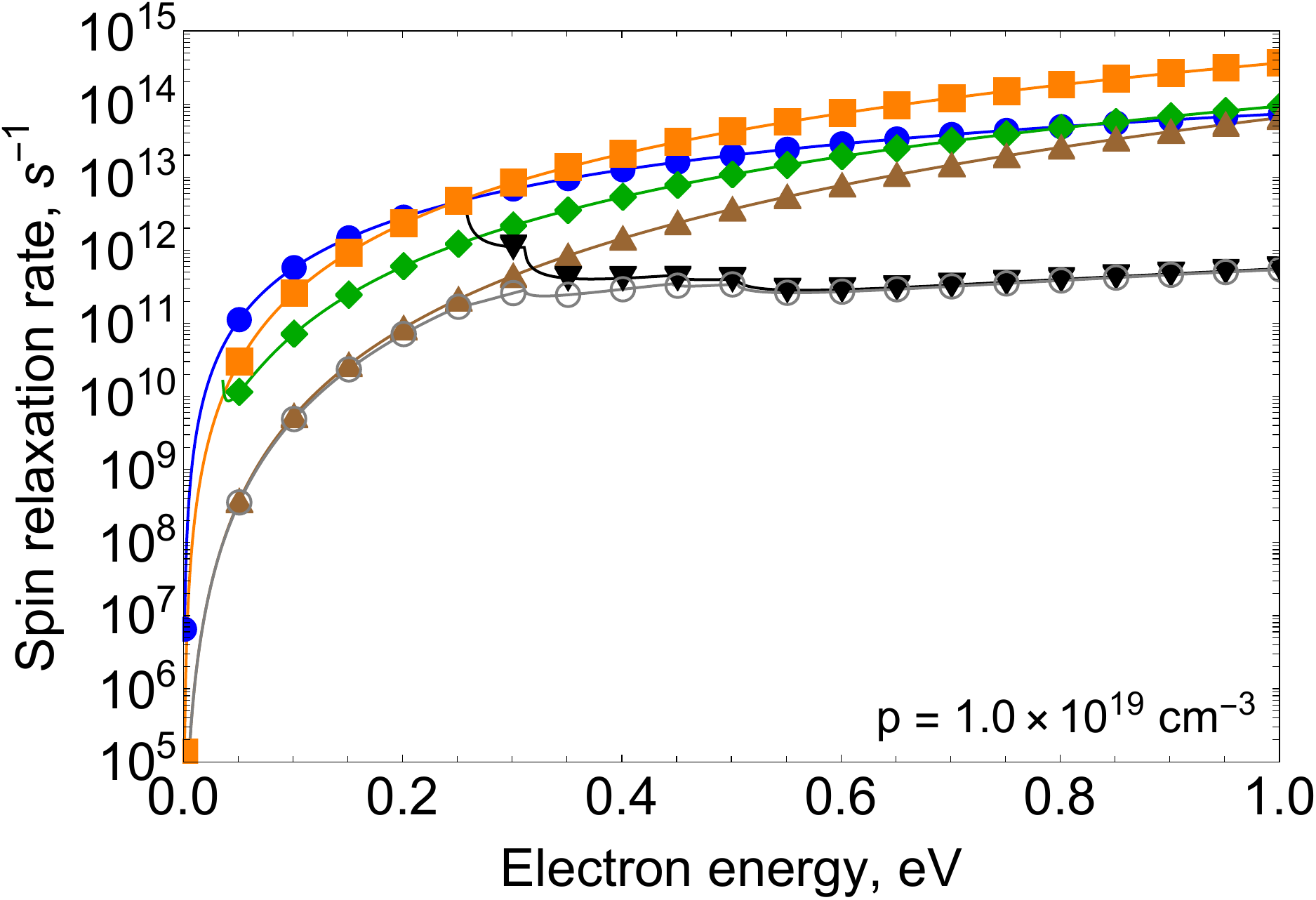}
\label{gamma_DP_19}}
\subfigure[][]{\includegraphics[width=3in]{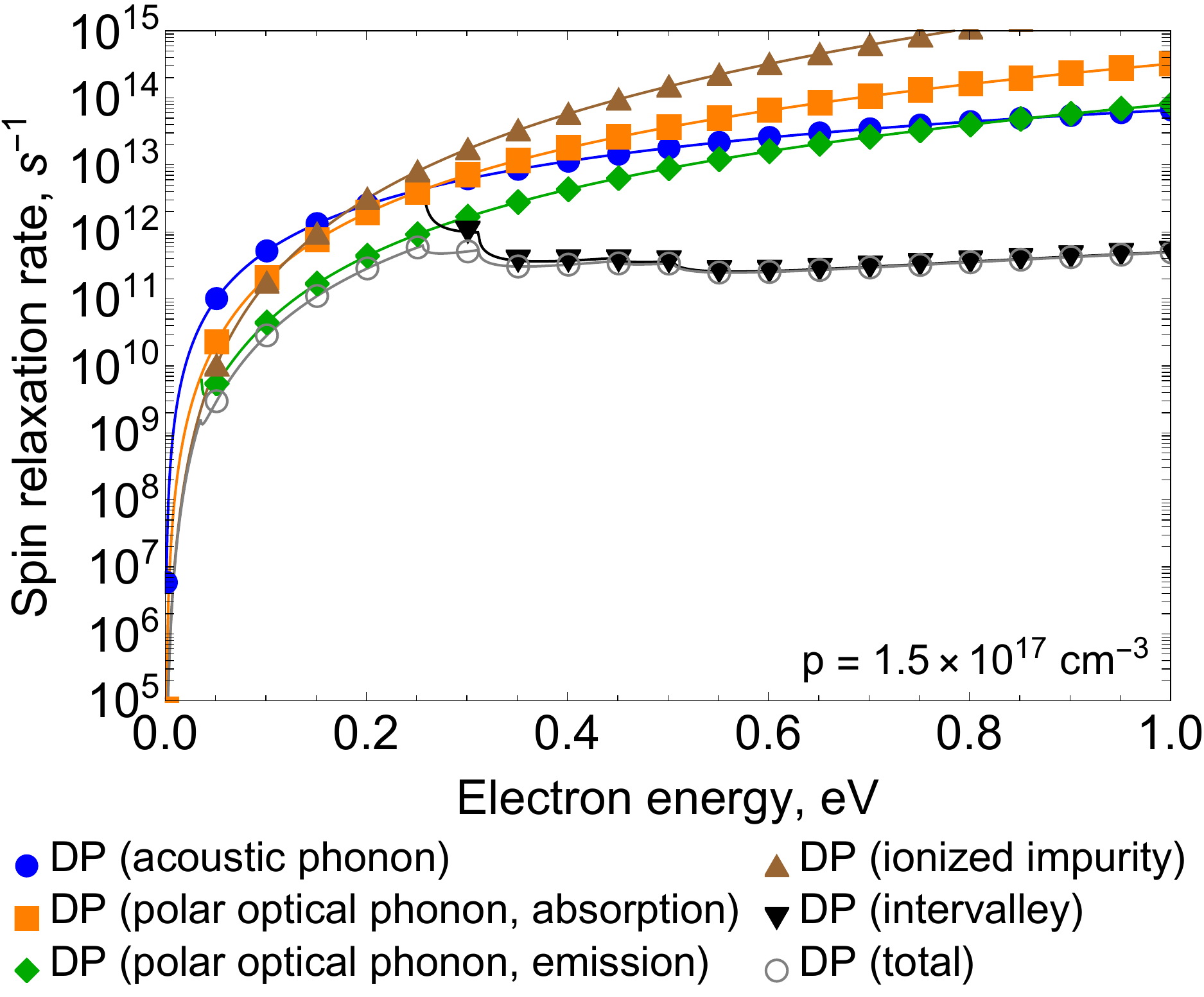}
\label{gamma_DP_17}}
\caption{D'yakonov-Perel spin relaxation rates due to different scattering mechanisms in the $\Gamma$ valley of (a) heavily doped and (b) moderately doped GaAs.}
\label{gamma_DP_19_17}
\end{figure}

\begin{figure}[!t]
\centering
\subfigure[][]{\includegraphics[width=3.in]{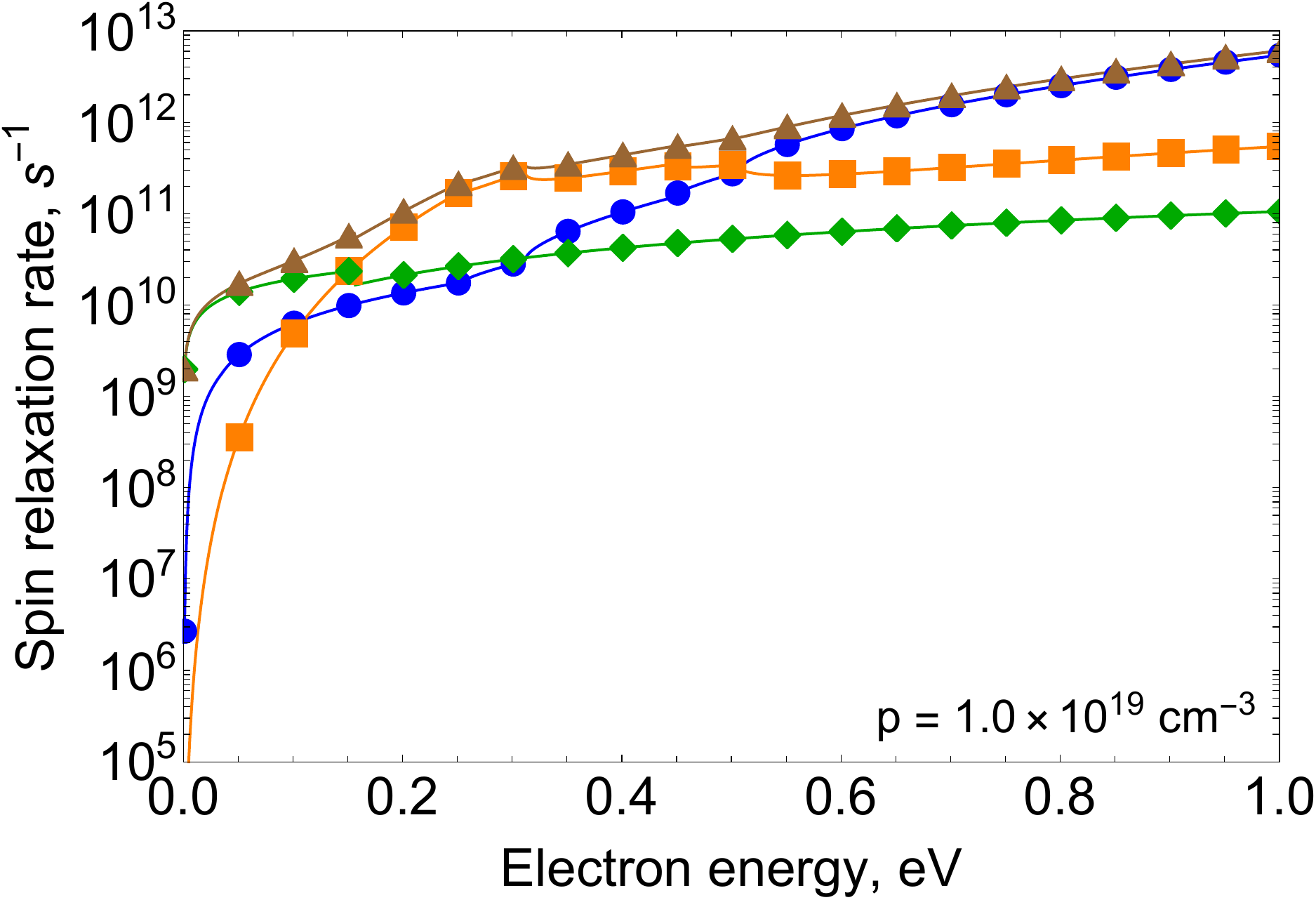}
\label{gamma_s_relax_19}}
\subfigure[][]{\includegraphics[width=3.in]{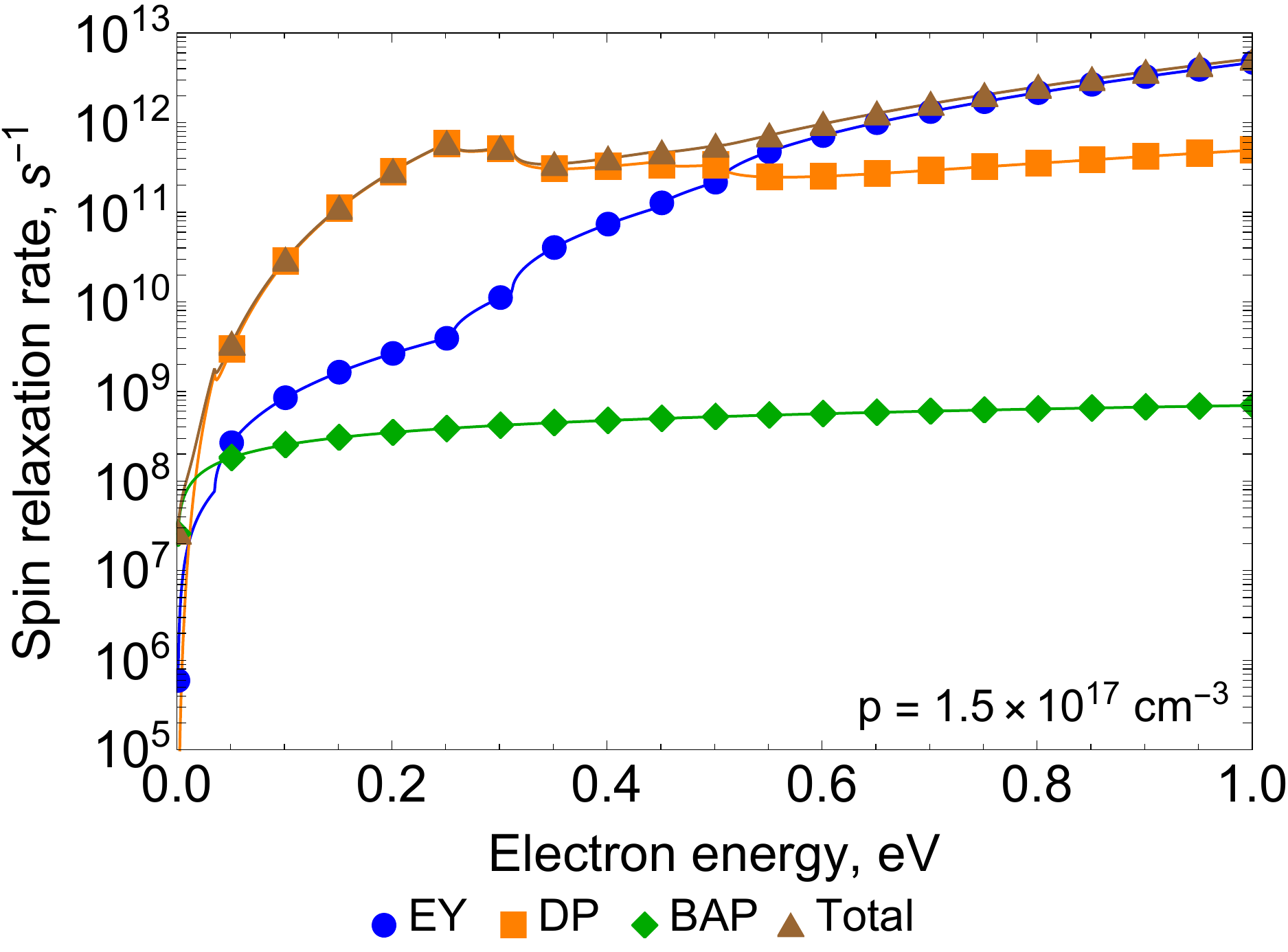}
\label{gamma_s_relax_17}}
\caption{Spin relaxation rates due to different mechanisms in the $\Gamma$ valley of (a) heavily doped and (b) moderately doped GaAs. Acoustic phonon scatterings, polar optical phonon scatterings, intervalley scatterings, and scatterings by ionized impurities were taken into account to calculate EY and DP spin relaxation rates.}
\label{gamma_s_relax_19_17}
\end{figure}

The strength of EY and DP spin relaxation mechanisms due to different scattering events in the $\Gamma$ valley is shown in Figs.~\ref{gamma_EY_19_17} and \ref{gamma_DP_19_17}, respectively, for two doping densities. In a moderately doped GaAs, interaction with polar optical phonons and ionized impurities are the dominant momentum relaxation mechanisms for the electrons whose energy does not exceed the energy required to be scattered into the upper valleys. The DP mechanisms is a main spin relaxation process in this case (Fig.~\ref{gamma_DP_17}). As the doping density increases, the DP mechanism affects the spin relaxation only through the interaction of high-energy electrons with impurities and the BAP mechanism becomes a dominant spin-relaxation process (Fig.~\ref{gamma_s_relax_19}). 

\begin{figure}[!b]
\centering
\subfigure{\includegraphics[width=3.in]{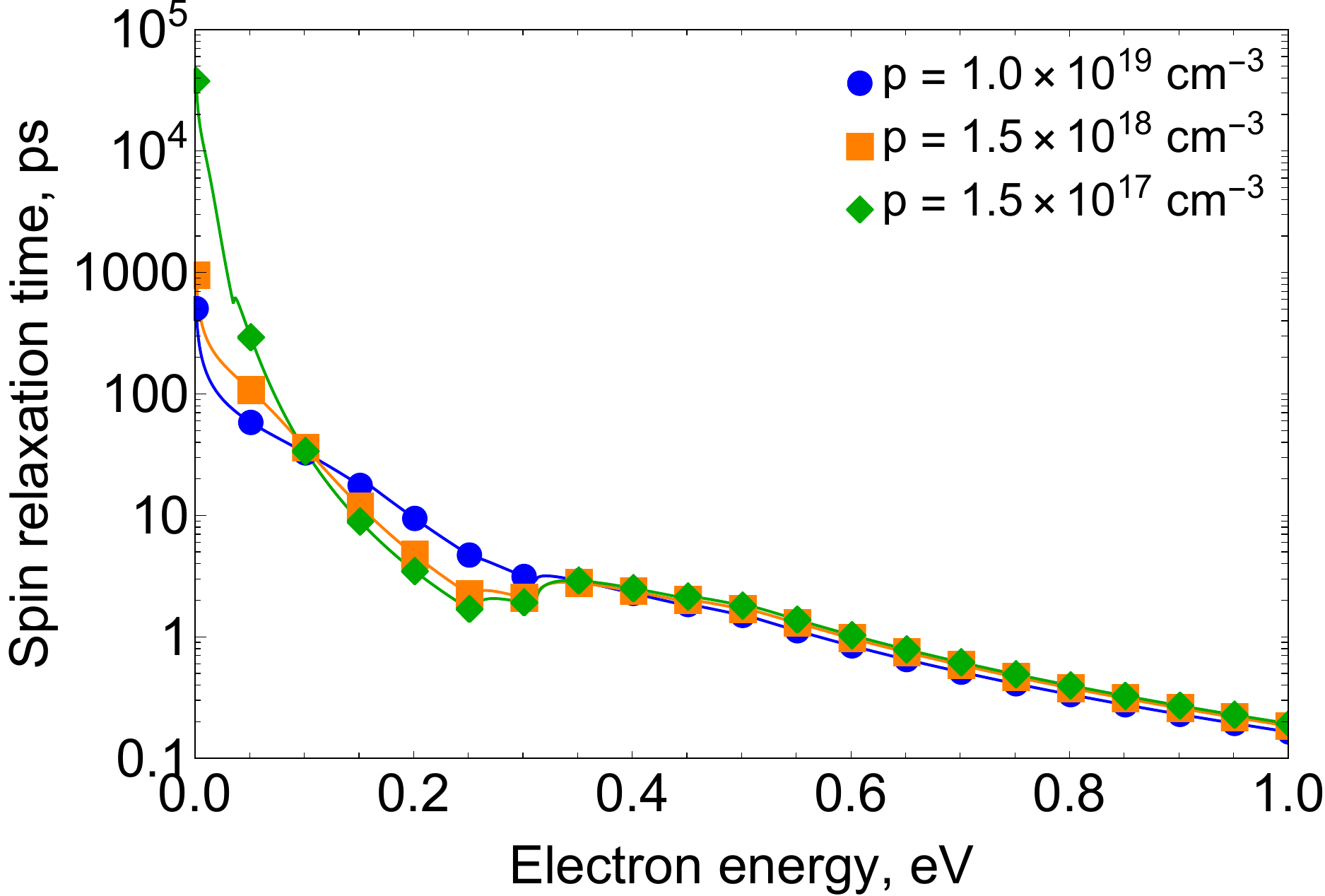}}
\caption{Spin relaxation time as a function of electron energy in the $\Gamma$ valley of GaAs calculated for three doping densities.}
\label{gamma_s_time_relax}
\end{figure}

Since the EY spin relaxation rate is proportional to the momentum relaxation rate, the total EY rate is simply given by the summation of contributions from all relevant scattering mechanisms. The DP spin relaxation rate is inversely proportional to the total momentum relaxation rate. Therefore, the total DP spin relaxation rate is calculated applying the Mathiessen's summation rule to the momentum relaxation rate\cite{Song_2002}
\begin{equation}
\frac{1}{\tau_\text{m}(\bold k)} = \frac{1}{\tau_\text{m}^{\text{ap}}(\bold k)} + \frac{1}{\tau_\text{m}^{\text{pop}}(\bold k)} + \frac{1}{\tau_\text{m}^{ij}(\bold k)} + \frac{1}{\tau_\text{m}^{\text{ii}}(\bold k)}.
\end{equation}
Then the total spin relaxation rate is given by
\begin{equation}
    \frac{1}{\tau_\text{s}(\bold k)} = \frac{1}{\tau_\text{s}^\text{EY}(\bold k)} + \frac{1}{\tau_\text{s}^\text{DP}(\bold k)} +\frac{1}{\tau_\text{s}^\text{BAP}(\bold k)}.
\end{equation}
The total spin relaxation rate $1/\tau_\text{s}$ and the spin relaxation time $\tau_\text{s}$ as a function of the electron energy in the $\Gamma$ valley are shown in Figs.~\ref{gamma_s_relax_19_17} and \ref{gamma_s_time_relax}, respectively. When the initial excess energy of photoexcited electrons is high, the spin relaxation time is much shorter as compared to that of thermalized electrons. However, the electrons quickly lose their energy in inelastic interactions with polar optical phonons and holes so the electron energy distribution stabilizes within 10 ps or so, even in the case of high initial excess energy. 

To implement the spin-relaxation mechanism in the Monte Carlo approach, we assume that the electron spin can flip in any scattering event. When the scattering happens, the random number $r$ is generated and if it is smaller than the spin-flip probability\footnote{Eq.~\ref{spin_flip_prob} can be easily understood if we rewrite the spin-flip probability $P$ in terms of $ESP$ and assume that $ESP$ decays exponentially with the decay time $\tau_\text{s}$, which gives \newline $P \equiv \frac{N_\downarrow}{N_\uparrow
 + N_\downarrow} = 0.5 \Big( \frac{N_\uparrow
 + N_\downarrow}{N_\uparrow
 + N_\downarrow} - \frac{N_\uparrow
 - N_\downarrow}{N_\uparrow
 + N_\downarrow}\Big) = 0.5 [1 - \exp(-\delta t/\tau_\text{s})].$}
\begin{equation}
    P(\bold k) = 0.5 \big[1 - \exp(-\delta t(\bold k)/\tau_\text{s}(\bold k))\big],
\label{spin_flip_prob}
\end{equation}
the spin changes its sign, otherwise the spin state is preserved. Here $\delta t$ is the time between any two consecutive scattering events excluding self-scatterings. In such formulation, $\tau_\text{s}$ also defines the polarization relaxation time, \ie the time required for the electron spin polarization to drop from its initial value $ESP_0$ to $1/e$, or 37$\%$, of that value.

\begin{figure}[!h]
\centering
\subfigure{\includegraphics[width=3.in]{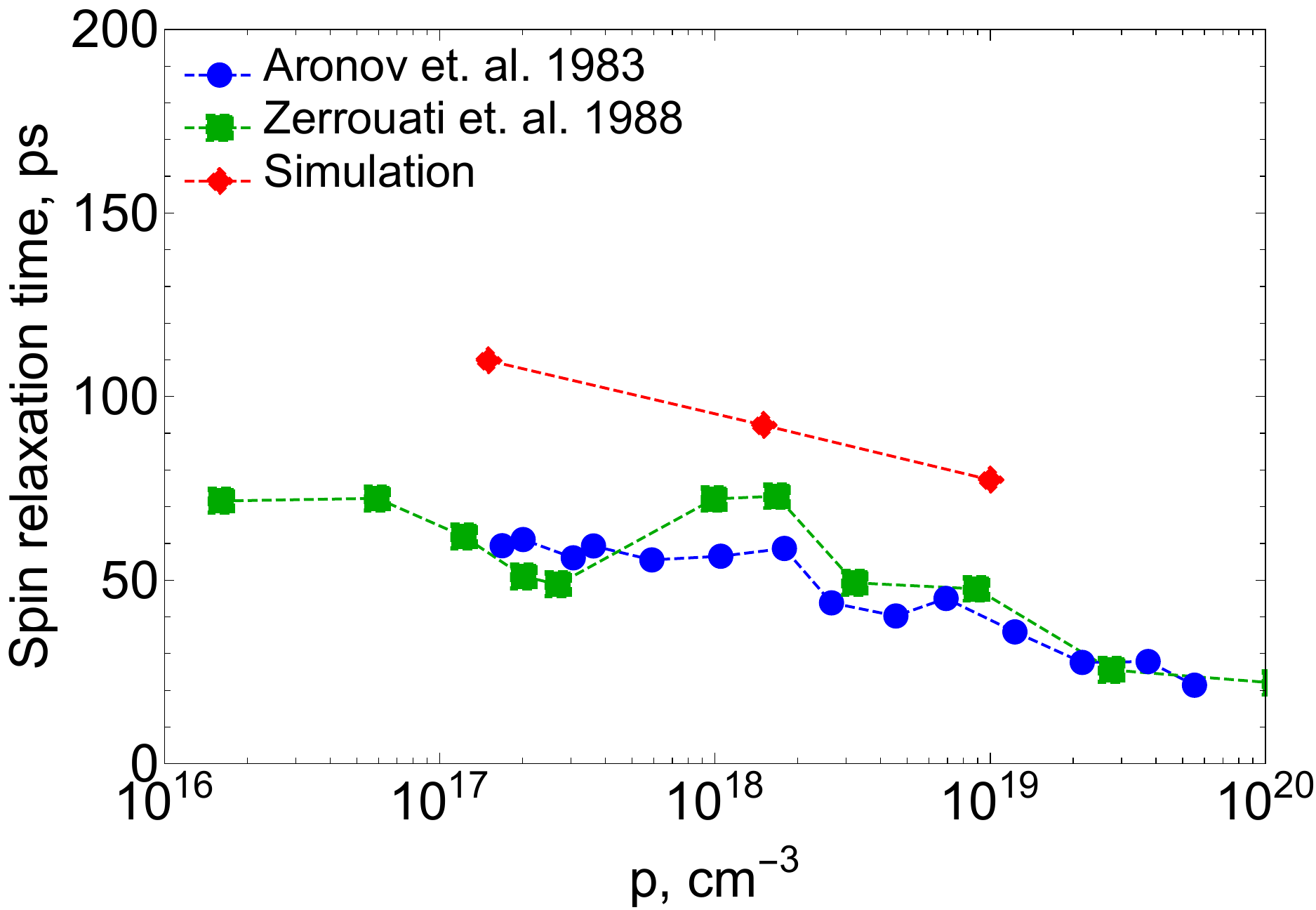}}
\caption{Spin relaxation time obtained from the decay of internal electron spin polarization for three doping densities and compared to available experimental data.\cite{Aronov_1983, Zerrouati_1988}}
\label{spin_rel_time}
\end{figure}

As it is shown above, the spin relaxation time $\tau_\text{s}$ depends on the electron kinetic energy, which dynamically changes in numerous scattering events. Obtaining the characteristic material spin relaxation time from the ensemble simulation is complicated mainly because of the way in which we implement the Pauli exclusion principle (theoretical calculations of energy-dependent spin relaxation time shown in Fig.~\ref{gamma_s_time_relax} do not account for the reduction of electron-hole interactions due to the degeneracy effect). However, it can be estimated from the decay of the internal electron spin polarization 
\begin{equation}
ESP^\text{int}(t) = \frac{N^\text{int}_{\uparrow}(t)-N^\text{int}_{\downarrow}(t)}{N^\text{int}_{\uparrow}(t)+N^\text{int}_{\downarrow}(t)},
\end{equation}
where $N^\text{int}_{\uparrow}$ and $N^\text{int}_{\downarrow}$ stand for the number of electrons inside the material with spin-up and spin-down states, respectively, calculated from the simulation when the emission is prohibited. In Fig.~\ref{spin_rel_time} we compare our results to the available experimental data\cite{Aronov_1983, Zerrouati_1988} where the electron spin relaxation time was estimated by measuring the polarisation of the luminescence from GaAs illuminated with a krypton-ion laser ($\hbar\omega = 1.65$ eV). The calculated spin relaxation time is 110 ps, 92~ps, and 77~ps for the doping density $1.5\times 10^{17}$~cm$^{-3}$, $1.5\times 10^{18}$~cm$^{-3}$, and $1\times 10^{19}$~cm$^{-3}$, respectively. Our results reproduce the same variation with doping density as the experimental data, however, are approximately factor of $1.4 - 1.8 $ larger. Despite this small discrepancy with the experimentally measured spin relaxation time, our simulations reproduce the measured spin polarization of the photoemitted electrons quite well (see Figs.~\ref{Paff} and \ref{Pdop} below) and even indicate that the calculated spin relaxation time is slightly underestimated.

\subsection{\label{sec:emission}Band Bending and Emission into the Vacuum}

\subsubsection{Band-Bending Region}

\begin{figure}[!b]
\centering
\subfigure[][]{\includegraphics[width=1.6in]{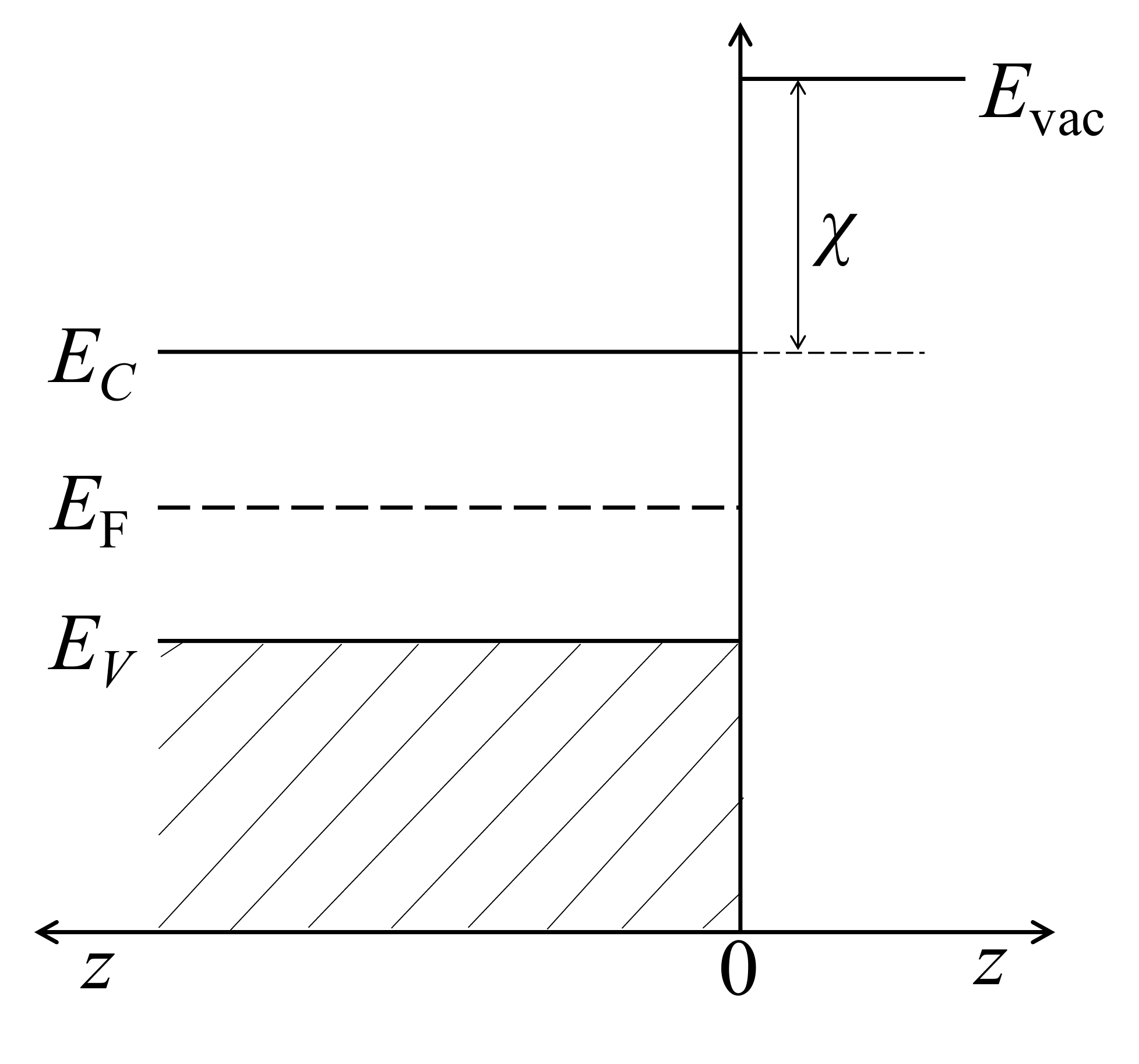}
\label{intrinsic}}
\subfigure[][]{\includegraphics[width=1.6in]{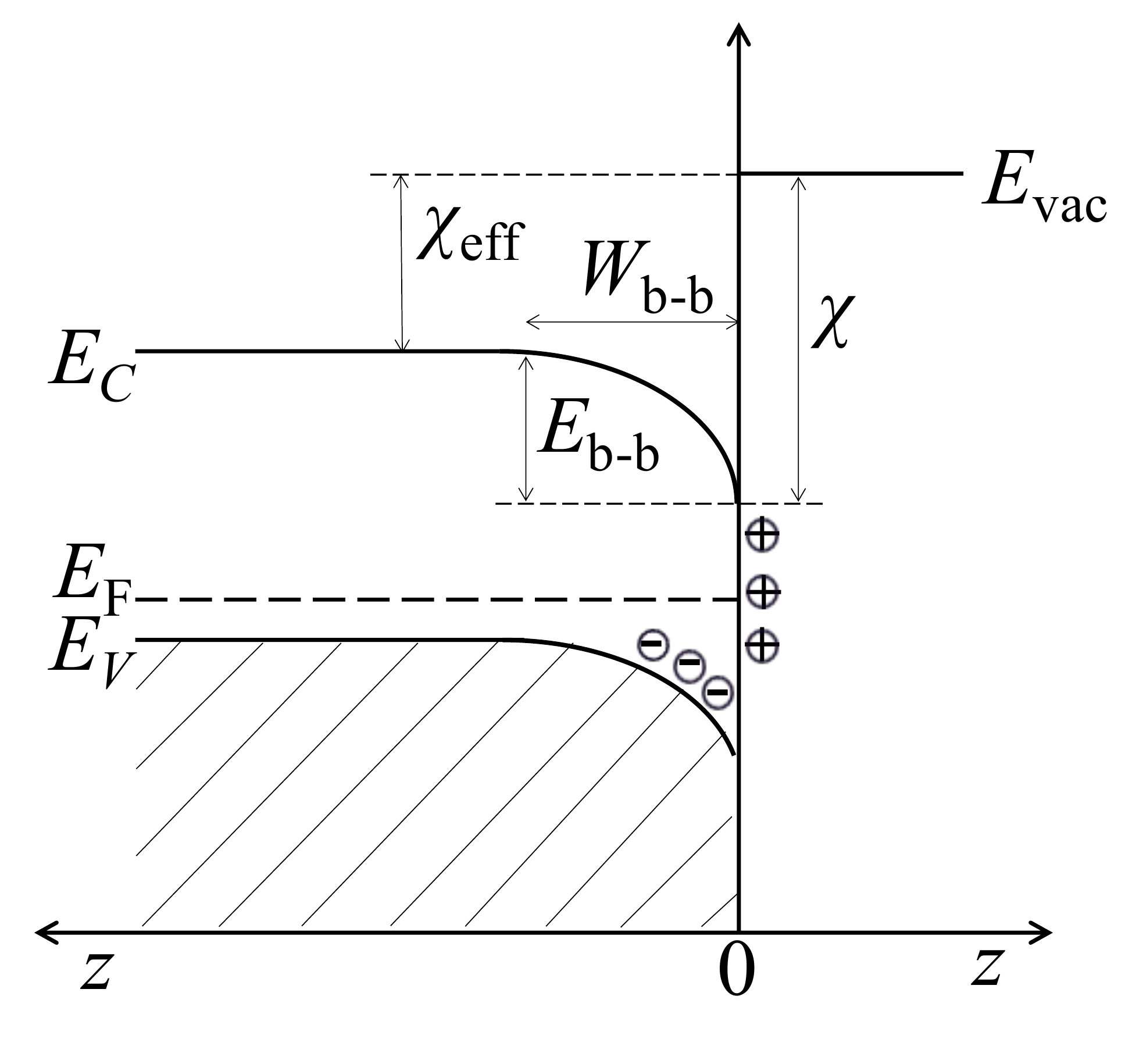}
\label{PEA}}
\subfigure[][]{\includegraphics[width=1.6in]{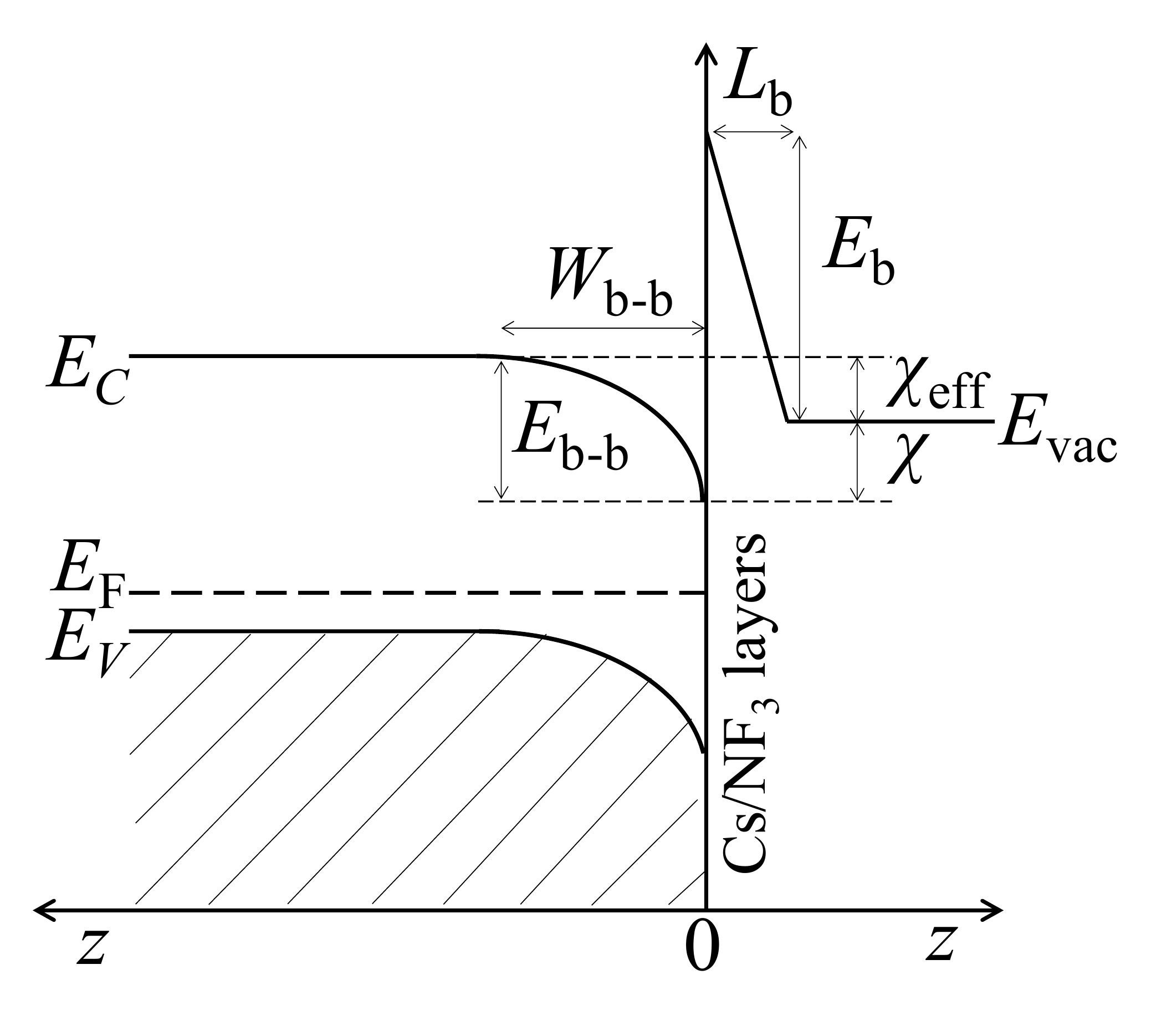}
\label{NEA}}
\caption{Effects of $p$-doping and surface coating on the threshold of photoemission from GaAs: (a) intrinsic PEA GaAs; (b) $p$-doped PEA GaAs; (c) $p$-doped NEA GaAs.}
\label{effects}
\end{figure}

An intrinsic GaAs has a Fermi level placed approximately in the middle of a band gap. Its clean surface is characterized by a high electron affinity $\chi \approx 4$ eV as shown in Fig. \ref{intrinsic}. In heavily $p$-doped semiconductors (Fig. \ref{PEA}), the Fermi level shifts toward the VB. In such materials, as discussed by Bell,\cite{Bell_Negative} electrons from surface states combine with acceptors in the VB leaving the surface positively charged. A positively charged surface on a $p$-type semiconductor repels the positively charged holes from the vicinity of the surface, forming a downward band-bending region and exposing a distribution of fixed negative charges (acceptor centers), in the volume of the semiconductor. This distributed charge leads to a variation of electric potential with depth which is equivalent to bending of the band edges. 

Electrons, photoexcited at a distance greater than the width of a band-bending region $W_{\text{b-b}}$ from the surface, are accelerated by the internal field with the result that they have to overcome only the "effective" electron affinity $\chi_{\text{eff}}$.\cite{Sommer_Photoemissive} To obtain the optimum enhancement of photoemission by means of band bending, it is required that both of the bands bend by the largest possible amount (which is $E_\text{g}$) in order to decrease the threshold of photoemission and that the bands bend within the shortest possible distance from the surface ($W_{\text{b-b}}$) since the threshold is maximally lowered only for those electrons that are photoexcited at a distance from the surface beyond the region of band bending.

\begin{figure}[!t]
\centering
\includegraphics[width=3.5in]{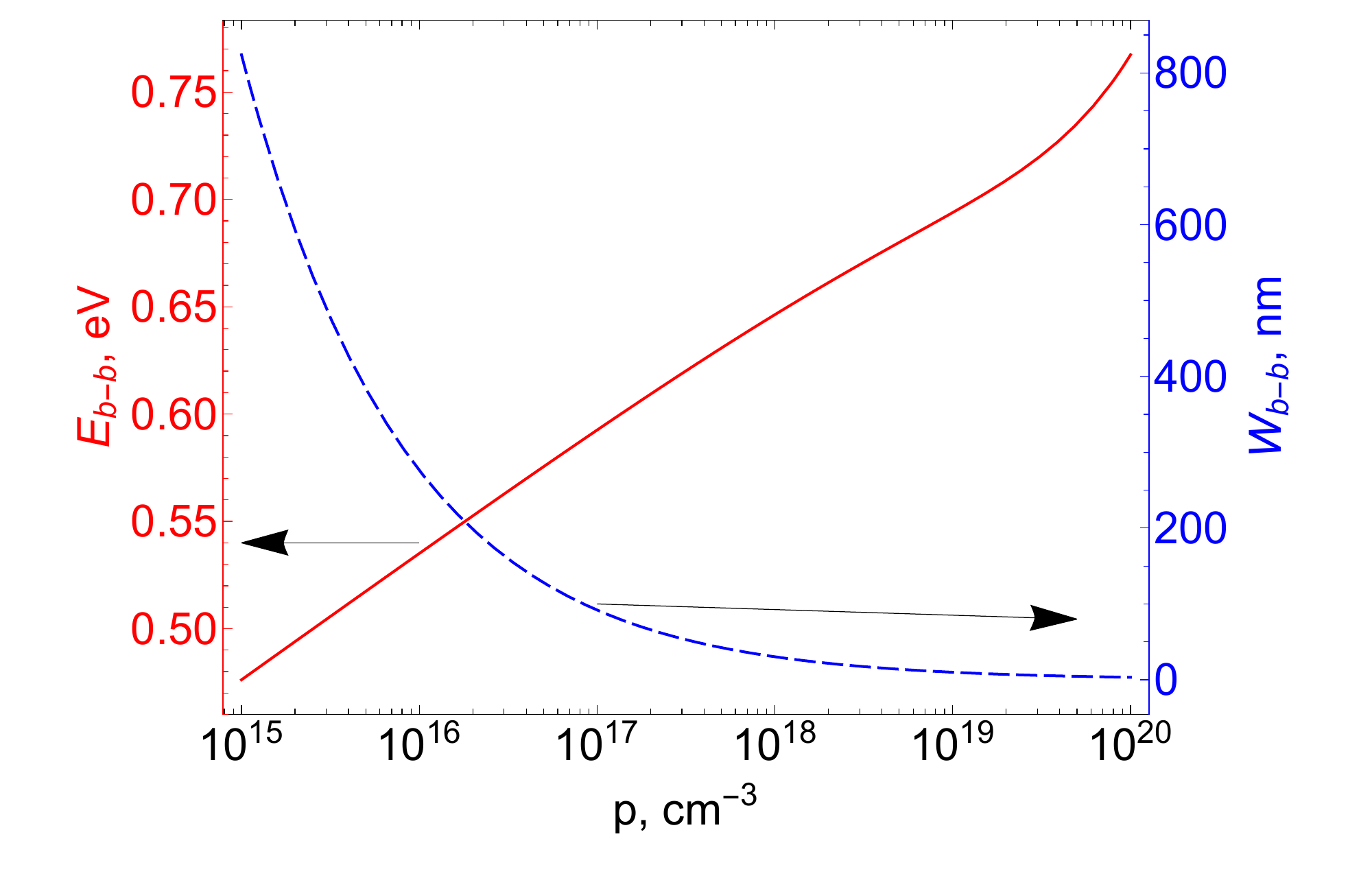}
\caption{Parameters of the band-bending region as a function of doping density. For $p=10^{19}$ cm$^{-3}$, $E_{\text{b-b}}= 0.694$ eV and $W_{\text{b-b}}= 9.947$ nm.}
\label{bb}
\end{figure}

The parameters of a band-bending region depend strongly on the doping density. The magnitude of the band bending $E_{\text{b-b}}$ in a $p$-type GaAs can be found as a difference between the Fermi level at the surface $E_\text{F}^\text{s}$ and the Fermi level in the bulk $E_\text{F}^\text{b}$\cite{Karkare_2013}
\begin{equation}
E_{\text{b-b}}=E_\text{F}^\text{s}-E_\text{F}^{\text{b}}.
\end{equation}
We assume that the Fermi level at the surface lies in the middle of the band gap
\begin{equation}
E_\text{F}^{\text{s}} \equiv E_\text{F}-E_\text{V}^{\text{s}}=\frac{1}{2}E_\text{g}.
\end{equation}
The position of the Fermi level relative to the VBM in the bulk can be calculated as\cite{Nilsson_1978}
\begin{equation}
\begin{split}
E_\text{F}^{\text{b}} &\equiv E_\text{F}-E_\text{V}^{\text{b}} \\
&= -k_\text{B}T\Bigg\{
\ln(p/N_\text{V})\\
&\quad +\frac{p/N_\text{V}}{\Big[64+0.05524\big(64+\sqrt{p/N_\text{V}}\big)p/N_\text{V} \Big]^{1/4}}
\Bigg\},
\end{split}
\end{equation}
where $N_\text{V}=2\big[ m_\text{h}^*k_\text{B}T/(2\pi \hbar^2)\big]^{3/2}$ is the effective density of states in the VB. This expression was also used to define the degeneracy condition for $p$-doped GaAs. We assume that the sample is degenerate if the Fermi level is as close to the VBM as 2$k_\text{B}T$ or lower. The width of a band-bending region is given by\cite{Stern_1972,Fisher_1972}
\begin{equation}
W_{\text{b-b}}=\sqrt{\frac{2\epsilon_\text{s}}{ep}|E_{\text{b-b}}|}.
\end{equation}   
The dependence of band-bending parameters on the doping density is shown in Fig. \ref{bb}.

\begin{figure}[!h]
\centering
\subfigure[][]{\includegraphics[width=1.6in]{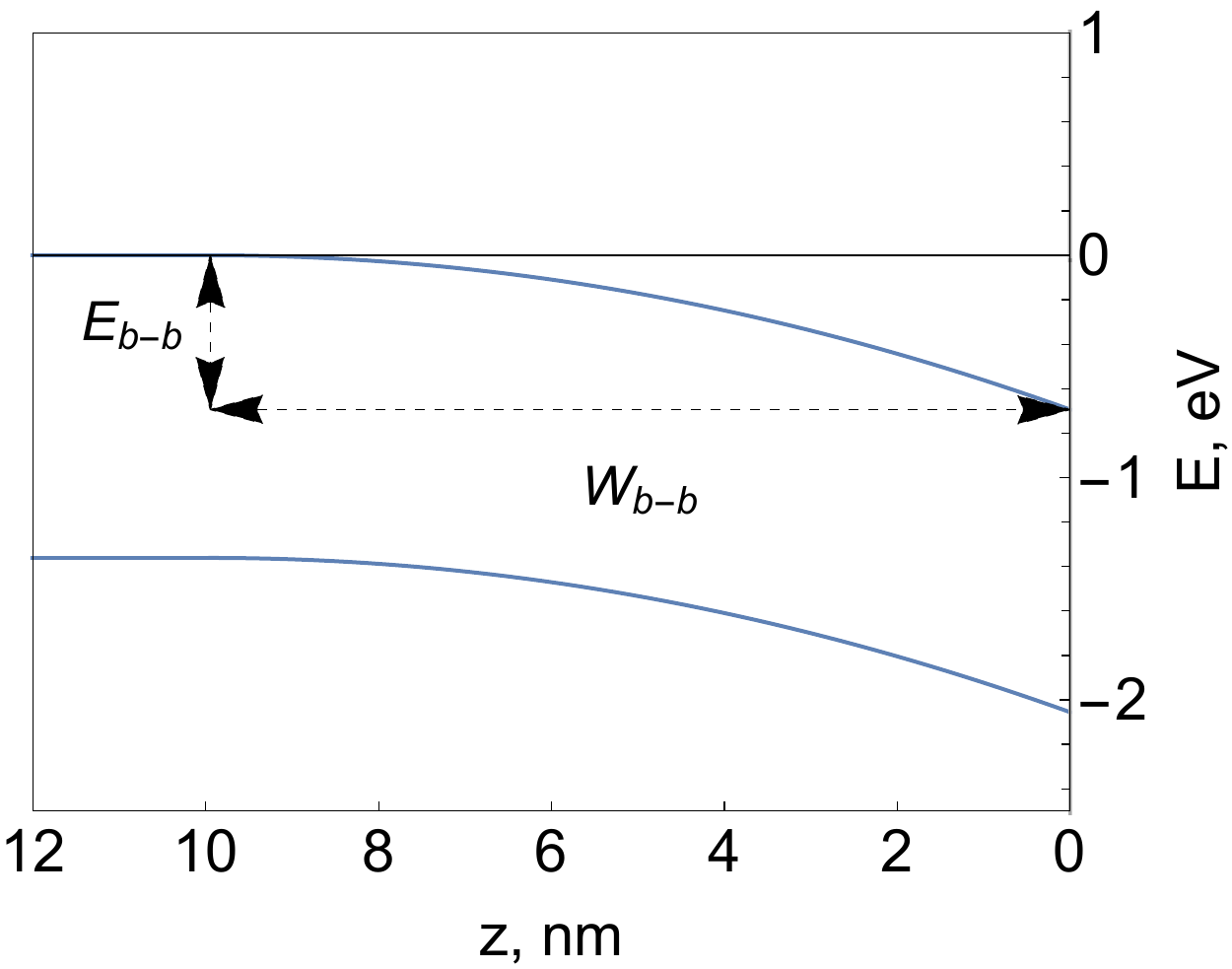}
\label{rect_pot_real}}
\hfil
\subfigure[][]{\includegraphics[width=1.65in]{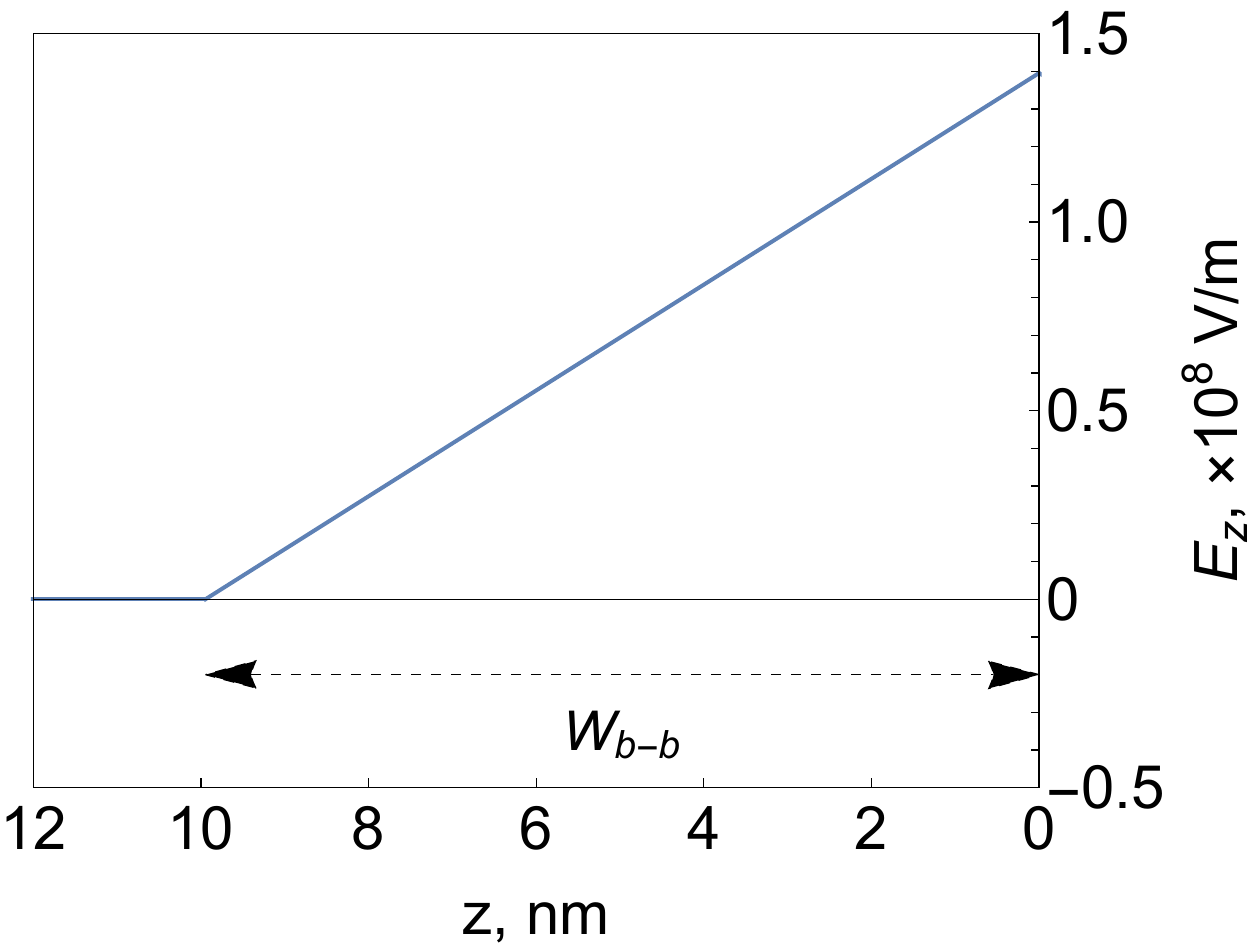}
\label{delta_field}}
\caption{(a) Real-size band-bending region for $p=10^{19}$ cm$^{-3}$ assuming the quadratic nature of bending given by Eq. \ref{d_Ec}. (b) Electric field that corresponds to the change of potential shown in (a).}
\label{EfEv_bb_real_size}
\end{figure}

We assume that the electric field in the bulk of the material is zero, \ie we only consider diffusion of electrons due to different scattering events. When electrons reach the band-bending region, they are exposed to the influence of the internal electric field $\mathbf{E}$ of the strength defined by the bending of band edges
\begin{equation}
\mathbf{E}(z) = -\frac{dV}{dz}=\frac{1}{e}\frac{dE_\text{C}}{dz}.
\end{equation}  
The exact value of the electric field as a function of the position of Fermi level at the surface of a semiconductor can be found by solving Poisson's equation. For simplicity, we assume that the bands bend according to the quadratic law
\begin{equation}
E_\text{C}(z) = \begin{cases}E_\text{C}^{\text{b}}, \qquad\qquad\qquad\qquad\quad z>W_{\text{b-b}}\\
E_\text{C}^{\text{b}}-E_{\text{b-b}}\big(1-\frac{z}{W_{\text{b-b}}}\big)^2, \quad 0<z<W_{\text{b-b}}
\end{cases}
\label{d_Ec}
\end{equation}
so electrons experience the influence of the electric field of the magnitude 
\begin{equation}
E_z = \begin{cases} 0, \qquad\qquad\qquad\quad\quad z>W_{\text{b-b}},\\
\frac{2E_{\text{b-b}}}{eW_{\text{b-b}}}\big(1-\frac{z}{W_{\text{b-b}}}\big), \quad 0<z<W_{\text{b-b}}.
\end{cases}
\end{equation} 
Numerical results are shown in Fig.~\ref{EfEv_bb_real_size}. We use a simple velocity Verlet algorithm\cite{VelocityVerlet} to implement the drift motion of electrons under the influence of this electric field. To ensure the energy conservation for electrons traveling through a narrow band-bending region in a heavily doped GaAs, we keep the simulation time step as short as 1 fs. 

\subsubsection{Surface Potential Barrier}

In addition to the band bending, $p$-doping in a semiconductor slightly reduces its electron affinity. The electron affinity can be reduced to much lower values by depositing thin layer of Cs. In general, electronic interaction between surface films and bulk materials causes the formation of a dipole which, depending on its orientation, reduces or increases the work function.\cite{Sommer_Photoemissive} Electropositive elements (\eg hydrogen, alkali and alkali earth metals) deposited on the material surface tend to decrease its work function. Alternating layers of Cs and NF$_3$ (or O$_2$) can effectively reduce the electron affinity to negative levels (Fig.~\ref{NEA}).

Electron affinity at the surface of Cs/NF$_3$-covered GaAs depends strongly on the details of the surface preparation and activation and its value is not known in the course of photoemission measurements. Moreover, comparison of theoretical calculations to experimental measurements indicates that the potential barrier exists at the surface of activated GaAs.\cite{Fisher_1972,Vergara_1999} The shape and size of the surface barrier determine how long electrons remain in the CB of a material before they escape or recombine. However, the exact shape and nature of this barrier are also not known. In this work, we use the triangular model of the surface potential barrier shown in Fig.~\ref{NEA} with $L_\text{b}=0.15$ nm and $E_{\text{b}}= 4$ eV, which was successfully used to explain behavior of photoemission characteristics from transmission mode NEA GaAs photocathodes.\cite{Vergara_1999} We use the electron affinity $\chi$ as a single fitting parameter. 

When the electron reaches the surface, it can scatter from the surface potential barrier, tunnel through it, or overcome it and be emitted into the vacuum. We use the propagation matrix method\cite{Levi_applied} to calculate the transmission probability and assume that the electron mass changes from $m_\text{e}^*$ to $m_0$ at the material-vacuum interface. Electrons photoexcited with high photon energies can be scattered into the $L$ and $X$ valleys. Moreover, electrons with small excess energy can be also scattered into the upper valleys in the band-bending region. However, escape only from the $\Gamma$ and some $X$ valleys is allowed by the momentum conservation law for the (100) GaAs surface.\cite{Vergara_1999, Karkare_2013} It should be noticed though that some papers\cite{Vergara_1996} report about the violation of the transversal momentum conservation at the real GaAs surface. This observation is attributed to the rough interface or a disordered activation layer, however the exact nature of this phenomenon is not known. In this work we assume that only electron emission from the $\Gamma$ and some $X$ valleys is allowed, \ie we assume an ideal surface.

\section{\label{sec:results}Results and Discussions}

The developed Monte Carlo program was used to calculate quantum efficiency and electron spin polarization simultaneously. In this section we compare our results with available experimental data.\cite{Chubenko_2014,Liu_2017} Assuming that the photoemission can be described by the Poisson statistics, the statistical error we should expect is the square root of the measured quantity (\ie if the number of emitted electrons is $N$, the statistical error is simply $\sqrt{N}$). Then the statistical error for QE and ESP is calculated according to the error propagation. To provide accurate statistics, $10^5$ particles were simulated for all calculations shown in this work. Electrons with the total energy below the vacuum level are trapped at the surface.\cite{Vergara_1999} Such electrons are removed from the simulation to reduce the computational time. The simulation time for all calculations was 370 ps, which corresponds to the recombination time in a heavily $p$-doped GaAs.\cite{Casey_1976} The recombination time in moderately doped GaAs is much longer (in a ns scale). However, the typical photocathode response time is much shorter than that.\cite{Bazarov_2008} So longer simulation time for moderately doped samples does not significantly affect results. Therefore, the accuracy of our results is mainly defined by the approximations introduced to derive certain expressions used in this work. 

\begin{figure}[!b]
\centering
\subfigure[][]{\includegraphics[width=3.in]{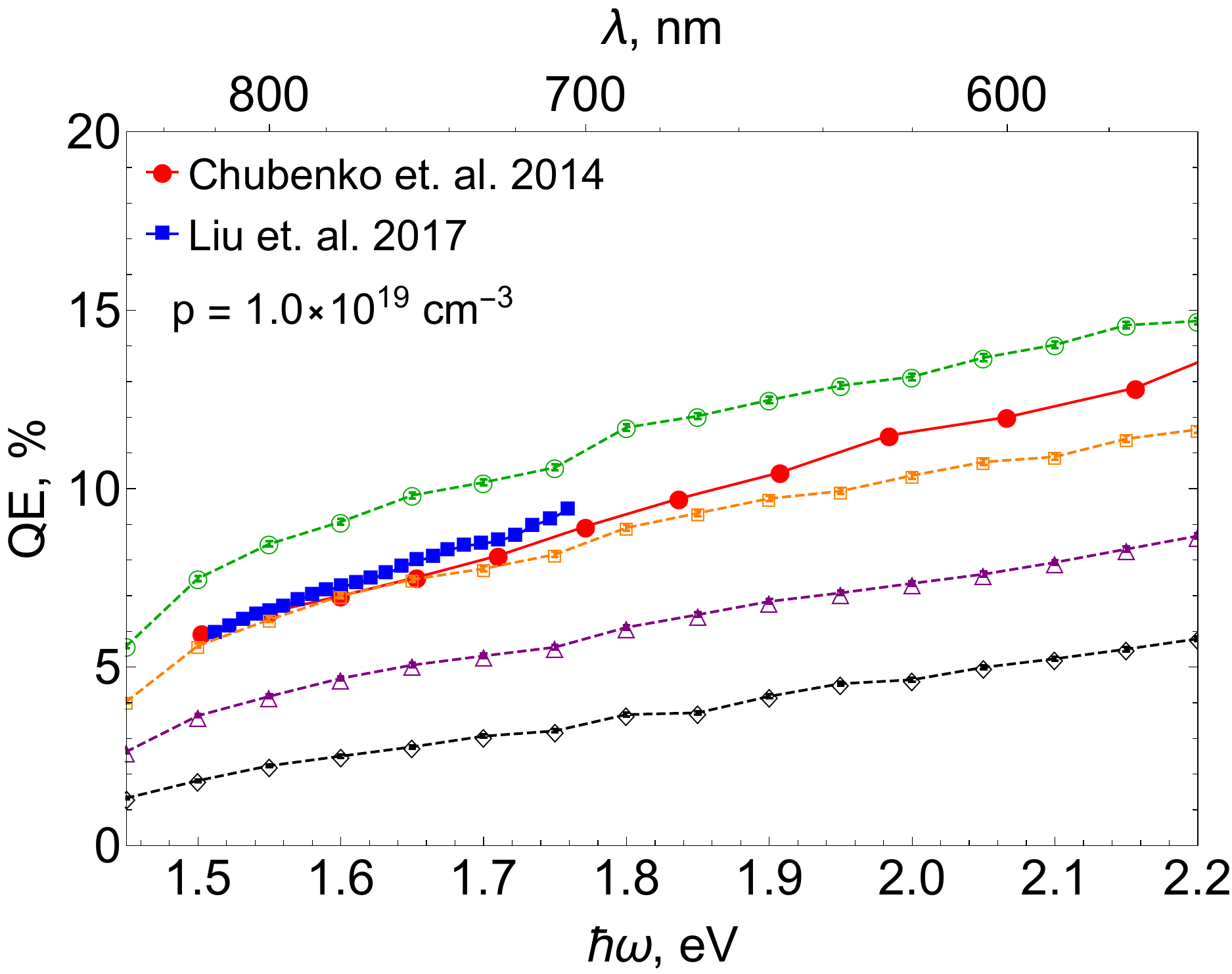}
\label{QEaff}}
\hfil
\subfigure[][]{\includegraphics[width=3.in]{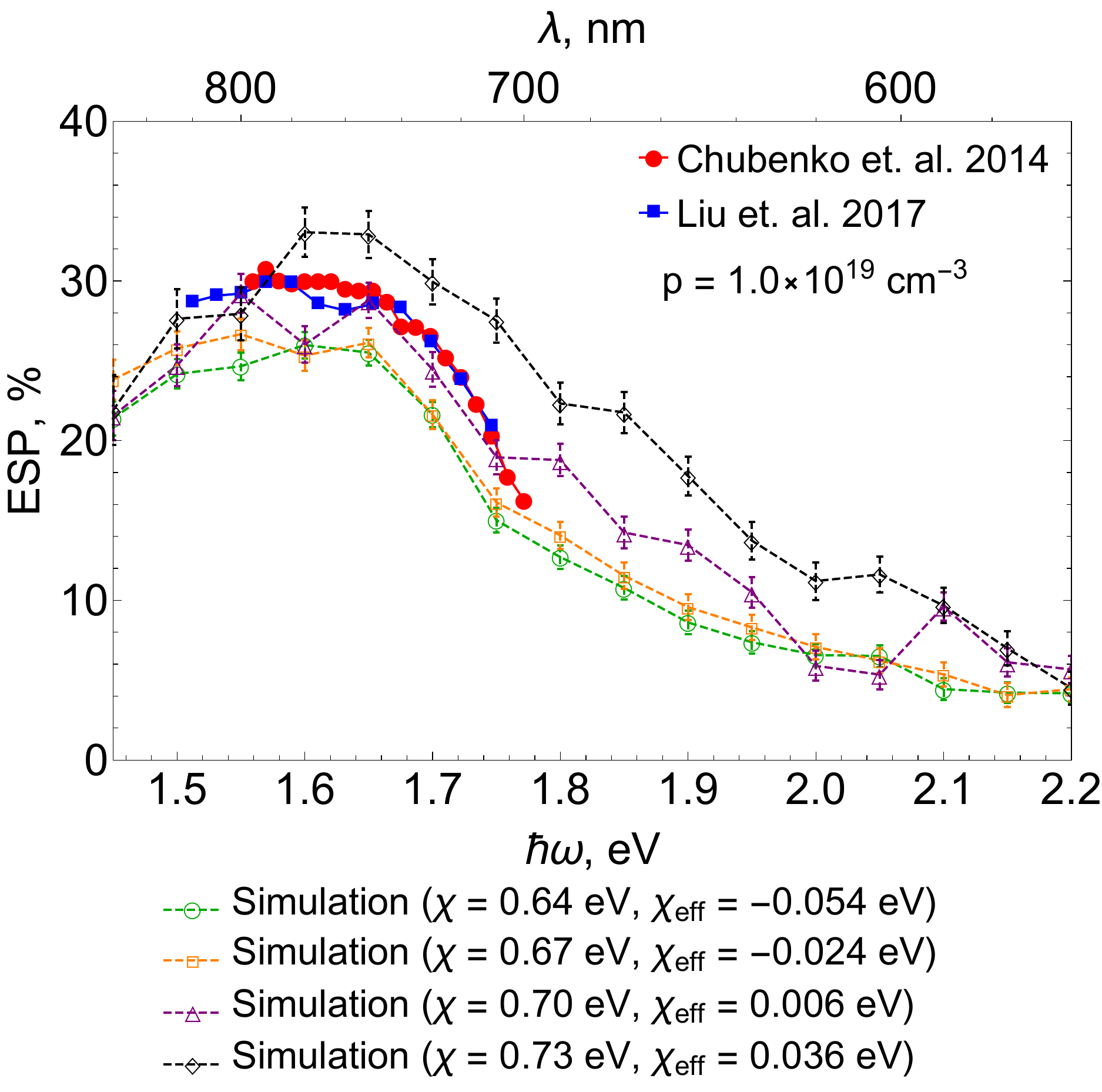}
\label{Paff}}
\caption{The calculated (a) QE and (b) ESP as a function of the photoexcitation energy for different electron affinity levels compared to the experimental data\cite{Chubenko_2014,Liu_2017} obtained from the sample with $p = 1 \times 10^{19}$ cm$^{-3}$ activated with O$_2$ and NF$_3$.}
\label{QEPaff}
\end{figure}

The results were obtained assuming the scattering mechanisms in the band-bending region to be identical to those in the bulk of the material. The electron affinity $\chi$ was used as a free model parameter. As it was discussed above, its value depends mostly on the quality of the activation procedure and remains unknown in the course of QE and ESP measurements. As can be seen in Fig.~\ref{QEPaff}, a good agreement with the experimental data can be achieved for both QE and ESP at $\chi=0.67$~eV, which corresponds to the negative effective electron affinity $\chi_{\text{eff}} = -0.024$~eV.

Explanation of the behaviour of QE curves in Fig.~\ref{QEaff} is very straightforward. Small-energy photons photoexcite electrons just above the bottom of the $\Gamma$ valley. Moreover, a great portion of such electrons is photoexcited at deep layers of the material. Therefore, small-energy electrons have smaller probability to reach the surface and overcome the surface potential barrier. In the case of photoexcitations by high-energy photons, many high-energy electrons exist close to the surface and have higher probability to be emitted into the vacuum. Therefore, the QE increases with photon energy. The QE also increases with decreasing the electron affinity level $\chi$, \ie with lowering the emission threshold.

As for the spin-polarization (Fig.~\ref{Paff}), low-energy electrons photoexcited with small-energy photons (larger than $E_\text{g}$, but smaller than $E_\text{g}+\Delta_\text{so}$) have high initial spin polarization $ESP_0$ and small spin-relaxation rate, resulting in the highest spin polarization. Great portion of electrons, photoexcited with a close-to-band-gap photon energy, exist deep in the material. These electrons undergo more scattering events during the diffusion motion from deep layers of the material towards the surface, which causes depolarization (more spin-flipping events). Therefore, the ESP is slightly smaller for photons with close-to-band-gap energies. The spin polarization drops fast when the emission from the $so$ sub-band becomes possible ($\hbar\omega > E_\text{g}+\Delta_\text{so}$). 

It is remarkable that a detailed behaviour of experimental ESP curves\cite{Liu_2017} as a function of the electron affinity can be reproduced in our calculations. Spin-polarization does not change significantly for negative and close-to-zero effective electron affinity levels. For large positive effective electron affinity, the ESP increases in a high photon-energy range and has a maximum at $\hbar \omega \approx 1.64$ eV, which is also observed in experiments (compare Fig.~\ref{Paff} below with the experimental ESP shown in Fig.~6 of Ref.~\onlinecite{Liu_2017}). Such behavior can be explained as follows. High electron affinity only allows emission of electrons with high excess energy. At high photoexcitation energies, electrons photoexcited from the $hh$ sub-band (spin-up state) are well energy separated from those photoexcited from $lh$ and $so$ sub-bands (spin-down states). Therefore, electrons with the same spin orientations are emitted during firs few ps after the photoexcitation, giving rise to the ESP. 

\begin{figure}[!t]
\centering
\subfigure[][]{\includegraphics[width=3.in]{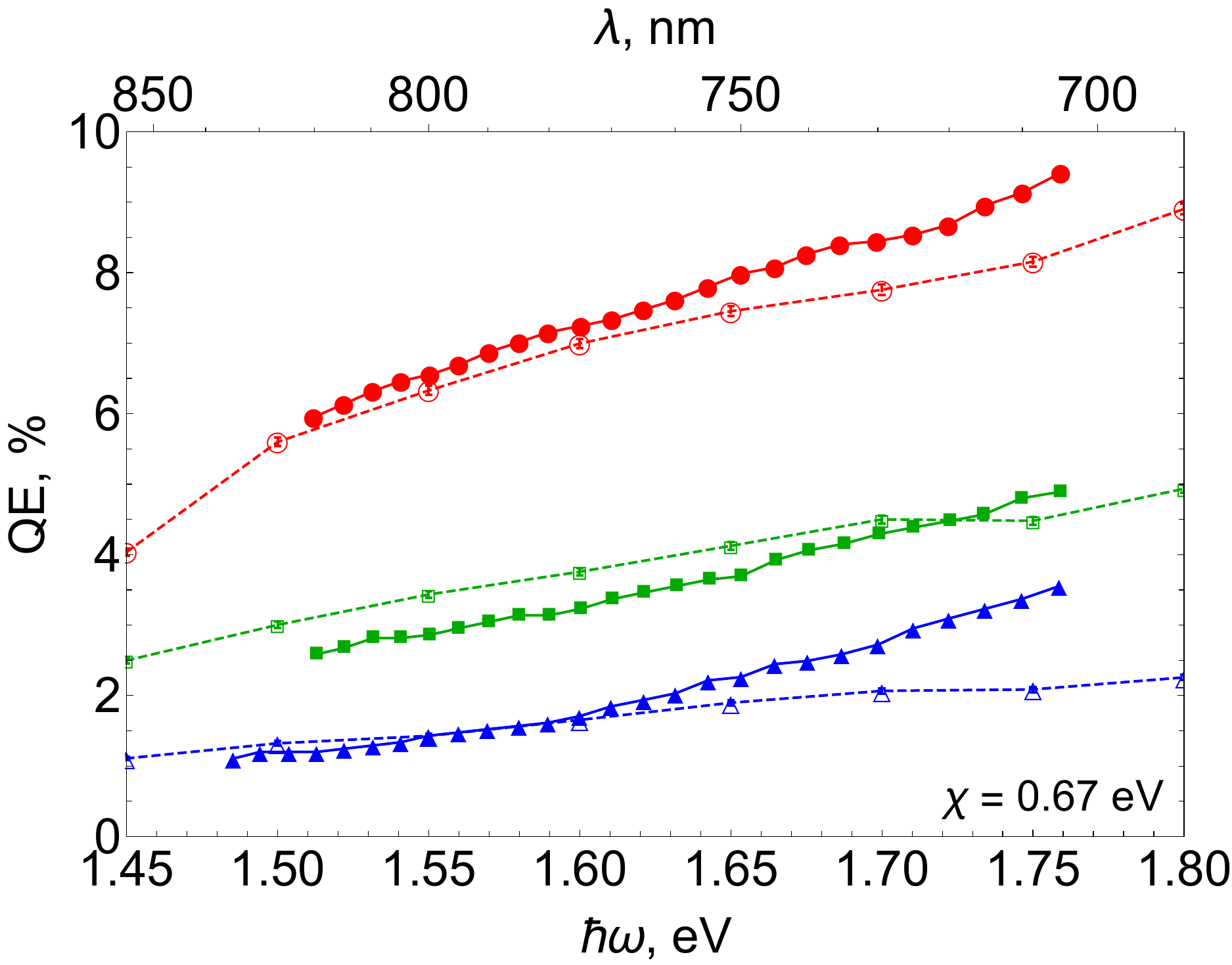}
\label{QEdop}}
\hfil
\subfigure[][]{\includegraphics[width=3.in]{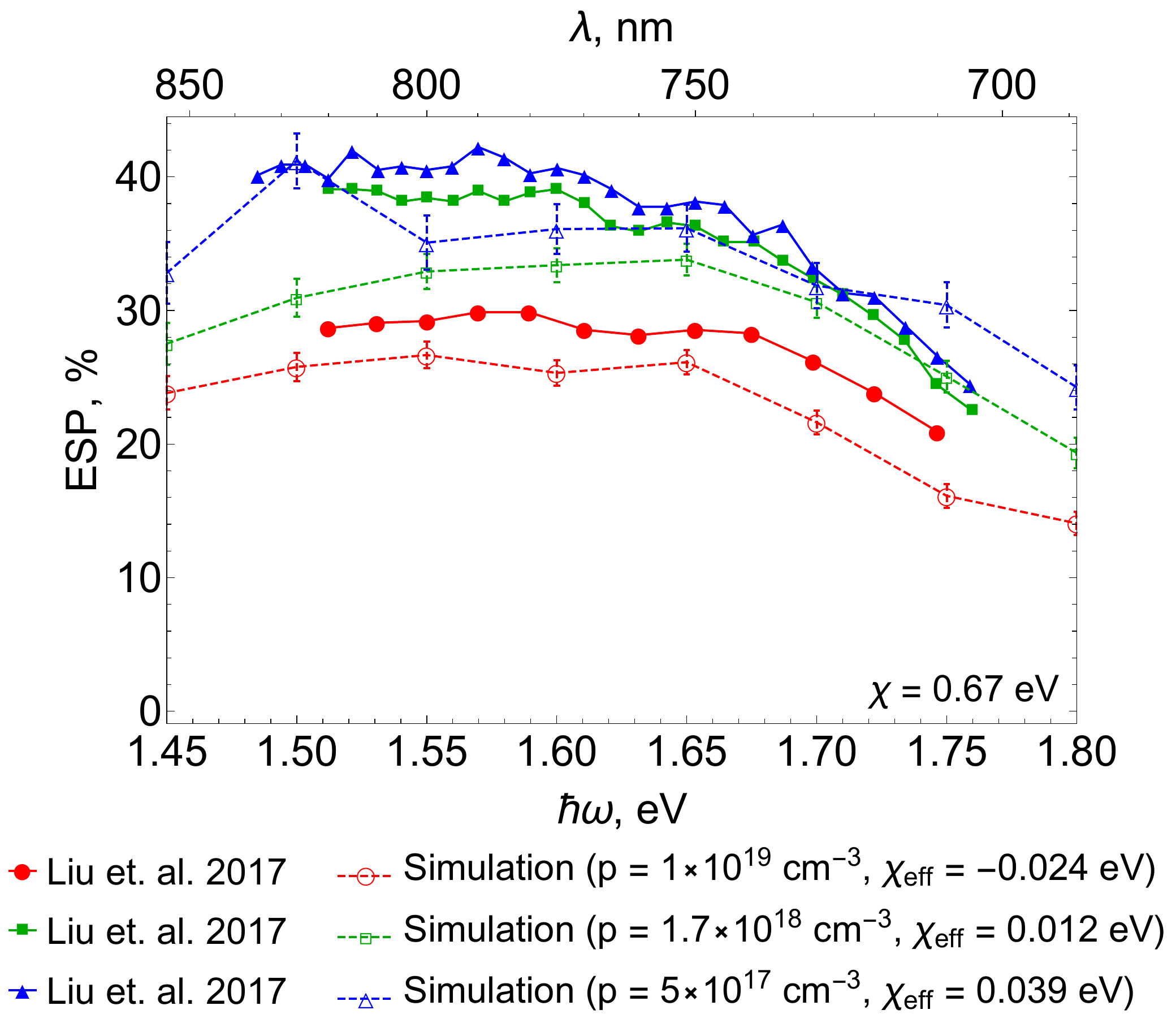}
\label{Pdop}}
\caption{Dependence of the (a) QE and (b) ESP on the doping density compared to the experimental data\cite{Liu_2017}. The electron affinity was fixed $\chi=0.67$ eV for all cases.}
\label{QEPdop}
\end{figure}

Using the electron affinity level determined for the sample with $p = 1\times 10^{19}$ cm$^{-3}$, the QE and ESP are calculated for different doping densities and are compared to the experimental data\cite{Liu_2017} in Fig.~\ref{QEPdop}. It means that we assume identical experimental conditions for all data sets. Even under this assumption, the agreement with the experimental data is pretty good for both QE and ESP. Better agreement for the QE can be obtained by slightly changing the electron affinity parameter for each experimental data set. 
 
\begin{figure}[!h]
\centering
\subfigure[][]{\includegraphics[width=3.in]{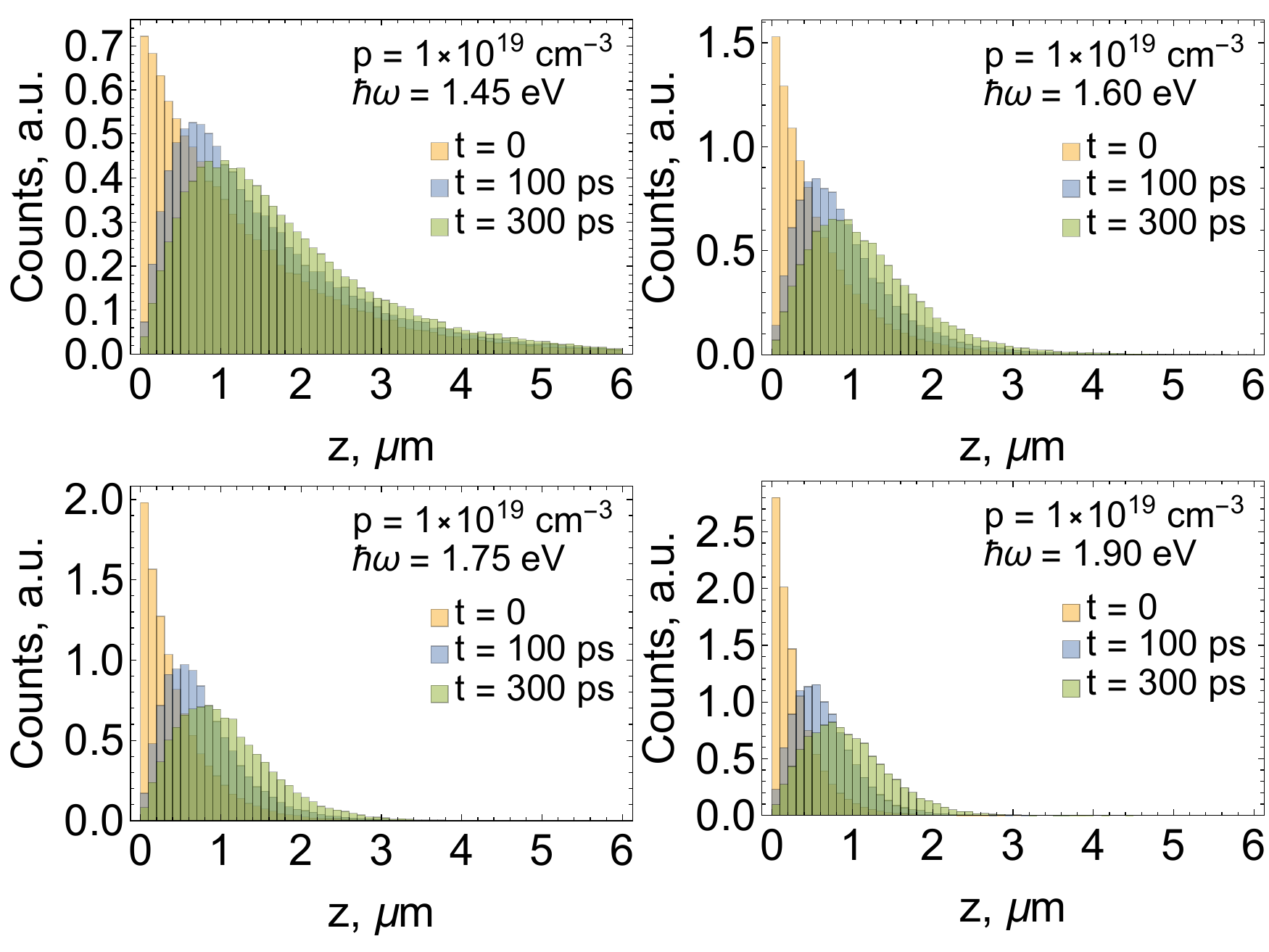}
\label{zf_19}}
\hfil
\subfigure[][]{\includegraphics[width=3.in]{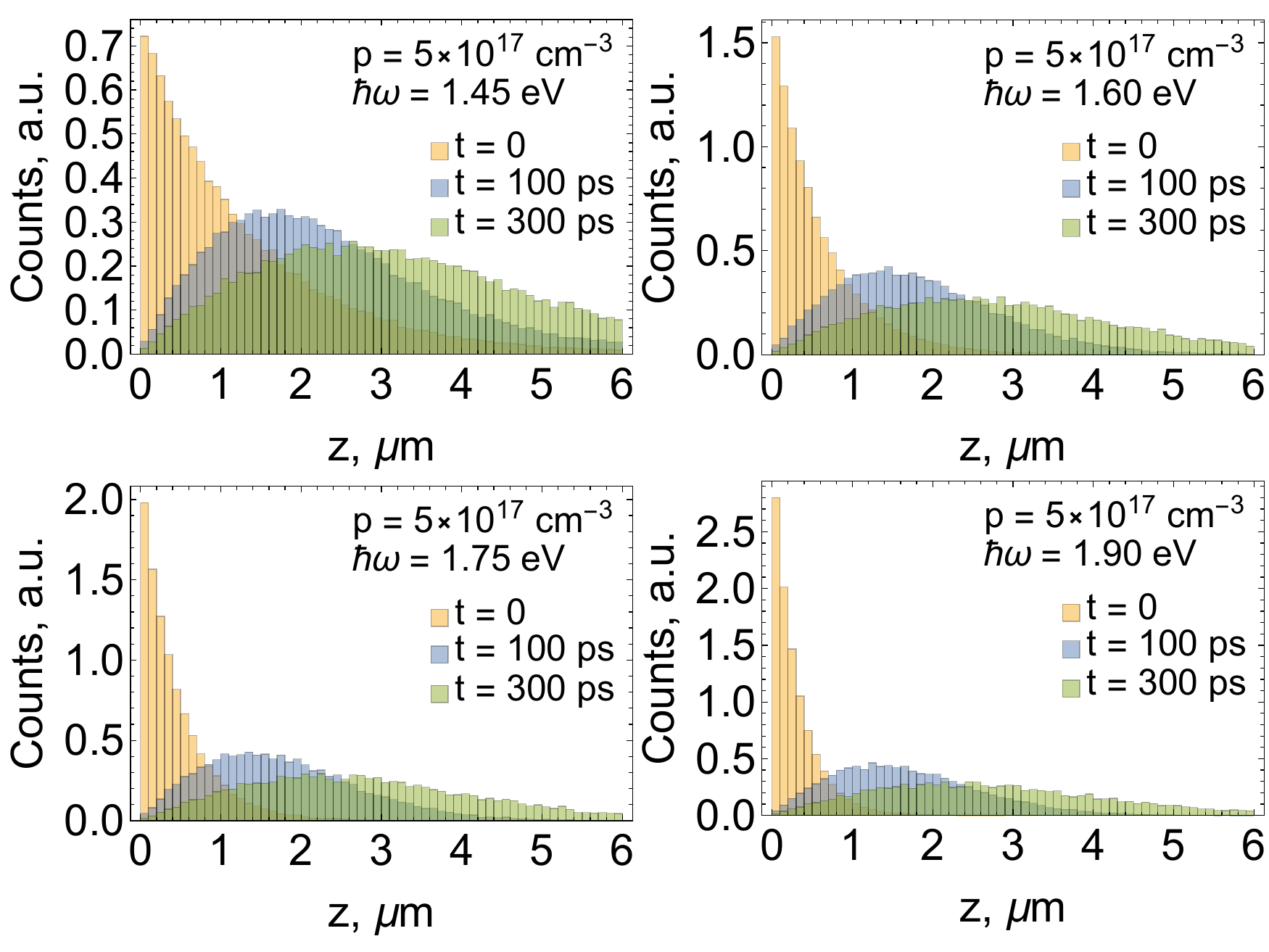}
\label{zf_17}}
\caption{Diffusion of not-emitted electrons in (a) heavily doped ($p = 10^{19}$ cm$^{-3}$) and (b) moderately doped ($p = 5\times 10^{17}$ cm$^{-3}$) GaAs.}
\label{zf}
\end{figure}

The highest QE can be obtained from the sample with the highest doping density. This can be achieved through an effective narrowing of the band-bending region (the band-bending region is as narrow as 10 nm for the sample with $p = 10^{19}$ cm$^{-3}$). Moreover, in heavily doped samples electrons stay longer at the surface because of the reduced diffusion process (see Fig.~\ref{zf}), which increases their probability to be emitted.

High doping density has an opposite effect for the ESP. The spin relaxation time is shorter for heavily doped samples. Therefore, electrons flip their spins faster while been close to the surface for a long time. Our ESP results are slightly smaller than the experimental measurements. This can be because we use a simple Brooks-Herring approach for the impurity momentum relaxation rate (Eq.~\ref{imp_rel_rate}), which does not correctly describe the momentum relaxation at high doping concentrations.\cite{Ridley_Quantum} Therefore, the DP spin relaxation mechanism is probably slightly overestimated. For more precise results, the approach which properly describes momentum relaxation rate for ionized impurity scattering in degenerate semiconductors should be used instead (see \eg Ref.~\onlinecite{Chung_1988}). We also assume that the electron can flip its spin in any scattering event independently on the leading spin-relaxation mechanism. It is known, however,\cite{Aronov_1983,Jiang_2009,Pikus_1984} that the DP mechanism causes the spin precession rather than the spin flip. Therefore, more accurate implementation\cite{Saikin_2003,Shen_2004} of the spin dynamics using the spin density matrix formalism\cite{Blum_Density_matrix, Kessler_Polarized_electrons, Bandyopadhyay_Introduction} would be beneficial.

\section{\label{sec:conclusions}Conclusions}
Using the Monte Carlo approach and details of the material band structure, the photoemission model has been implemented and used to calculate QE and ESP from GaAs photocathodes. Both momentum and spin relaxation rates due to the scattering of electrons on acoustic phonons, optical (polar and non-polar) phonons, ionized impurities, and the interactions with holes are calculated as a function of electron energy for various doping densities. We employ a simple triangular model of the surface potential barrier and use the electron affinity $\chi$ as a fitting parameter to compare calculation results with experimental data for NEA GaAs with $p = 10^{19}$ cm$^{-3}$.

It is remarkable that the same value of parameter $\chi$ can be used to achieve a good agreement with experimental QE for different doping densities without a need of making any further assumptions or modifications in the code. We find that the rate, at which electrons change their spins, is defined by the interactions with holes through the BAP mechanism in heavily doped samples. In moderately doped GaAs, the DP mechanism through interactions with polar optical phonons is the most effective spin-relaxation mechanism. With all the assumptions and approximations used in this work, we obtain a good agreement with experimental data available in literature. However, more accurate description of the DP spin relaxation mechanism and ionized impurity scatterings would be beneficial.   

We conclude that the QE and ESP of electrons photoemitted from GaAs at room temperature can be fully explained by the bulk relaxation mechanisms and the time which electrons spend in the material before being emitted. The developed model can be used as an effective scientific tool to study momentum and spin relaxation mechanisms as well as photoemission characteristics of other spin-polarized electron sources based on III-V family semiconductors. Moreover, the Monte Carlo code can be easily modified to study QE and ESP not only from bulk samples, but also from more complex geometries (\eg thin photocathodes, layered superlattices, \etc). The model is also essential for deeper understanding of other photocathode parameters such as mean transverse energy of photoemitted electrons, photocathode response time, and surface properties of activated photocathodes.

\begin{acknowledgments}
This work was supported by The George Washington University in the form of a graduate student fellowship for O.C. Some authors (O.C., S.K., J.K.B., and I.B.) were also supported by the U.S. National Science Foundation under Award PHY-1549132, the Center for Bright Beams. 

The authors are grateful to Dr.~Ievgen Lavrukhin for his help with code optimization. We also acknowledge Research Computing at Arizona State University for providing HPC resources.
\end{acknowledgments}

\section*{Data Availability}

The data that support the findings of this study are available from the corresponding author upon reasonable request.

\bibliography{References}

\end{document}